\documentclass[journal,10pt,draftclsnofoot,onecolumn]{IEEEtran}
\usepackage{amsmath,amsfonts,amssymb}
\usepackage{multirow,float,makecell}
\usepackage{hyperref}
\usepackage{amsthm}
\usepackage[numbers]{natbib}
\usepackage{comment}
\newcommand{\conKp}{K}
\newcommand{\conNp}{N}

\newcommand{\dldim}{6}
\newcommand{\dldimt}{9}

\newcommand{\getsr}{\leftarrow_{\textsc r}}

\newcommand{\Remark}{\noindent {\em Remark:} }

\newcommand{\Z}{\mathbb{Z}}
\newcommand{\mdp}{\textsf{\textup{mod~}}}

\newcommand{\Bdual}{\mathbb{B}}
\newcommand{\Ddual}{\mathbb{D}}
\newcommand{\Fdual}{\mathbb{F}}
\newcommand{\Dual}{\textsf{\textup{Dual}}}

\newcommand{\DDHop}{\textsf{\textup{DDH1}}}
\newcommand{\DDHtp}{\textsf{\textup{DDH2}}}
\newcommand{\DSop}{\textsf{\textup{DS1}}}
\newcommand{\DStp}{\textsf{\textup{DS2}}}

\newcommand{\DLINp}{\textsf{\textup{DLIN}}}

\newcommand{\bigO}{\mathcal{O}}

\newcommand{\AdvA}{\mathcal{A}}
\newcommand{\AdvB}{\mathcal{B}}

\newcommand{\Adv}{\mathsf{Adv}}

\newcommand{\SFp}{\textsf{\textup{SF}}}

\newcommand{\Rp}{\textsf{\textup{R}}}

\newcommand{\Gm}{\mathsf{Game}}

\newcommand{\qnop}{q_{n_1}}
\newcommand{\qntp}{q_{n_2}}

\newcommand{\SimSetup}{\textsf{\textup{Setup}}}
\newcommand{\SimQuery}{\textsf{\textup{Query}}}

\newcommand{\SimChallenge}{\textsf{\textup{Challenge}}}
\newcommand{\SimGuess}{\textsf{\textup{Guess}}}

\newcommand{\msgp}{\mathrm{m}}
\newcommand{\idp}{\mathrm{id}}

\newcommand{\MK}{\mathsf{MK}}
\newcommand{\PP}{\mathsf{PP}}

\newcommand{\CT}{\mathsf{CT}}
\newcommand{\SK}{\mathsf{SK}}
\newcommand{\sKp}{\mathsf{K}}
\newcommand{\sCp}{\mathsf{C}}

\newcommand{\Setup}{\textsf{\textup{Setup}}}

\newcommand{\Enc}{\textsf{\textup{Enc}}}
\newcommand{\Dec}{\textsf{\textup{Dec}}}

\newcommand{\RIBEp}{\textsf{\textup{RIBE}}}

\newcommand{\matA}{\mathbf{A}}

\newcommand{\veczero}{\mathbf{0}}

\newcommand{\vecb}{\mathbf{b}}
\newcommand{\vecd}{\mathbf{d}}
\newcommand{\vecf}{\mathbf{f}}

\newcommand{\vecv}{\mathbf{v}}
\newcommand{\vecw}{\mathbf{w}}

\newcommand{\vecx}{\mathbf{x}}

\newcommand{\vecz}{\mathbf{z}}

\newcommand{\tran}{{\!\scriptscriptstyle{\top}}}

\newcommand{\PriKeyGen}{\textsf{\textup{PriKeyGen}}}
\newcommand{\DecKeyGen}{\textsf{\textup{DecKeyGen}}}
\newcommand{\KeyUpd}{\textsf{\textup{KeyUpd}}}
\newcommand{\KeyRev}{\textsf{\textup{KeyRev}}}

\newcommand{\Msgsp}{\mathcal{M}}
\newcommand{\Idsp}{\mathcal{I}}
\newcommand{\Tmsp}{\mathcal{T}}

\newcommand{\KU}{\mathsf{KU}}
\newcommand{\DK}{\mathsf{DK}}

\newcommand{\tp}{\mathrm{t}}

\newcommand{\RLp}{\mathsf{RL}}
\newcommand{\STp}{\mathsf{ST}}

\newcommand{\Path}{\mathsf{Path}}
\newcommand{\BT}{\mathsf{BT}}
\newcommand{\Root}{\mathsf{root}}

\newcommand{\KUNodes}{\textsf{\textup{KUNodes}}}

\begin{document}
\title{Anonymous and Adaptively Secure Revocable IBE with Constant-Size Public Parameters}

\author{Jie Chen$^\ast$\thanks{$^\ast$Corresponding Author},
        Hoon Wei Lim,
        San Ling,
        Le Su
        and~Huaxiong~Wang%

\thanks{$^{\ast\ast}$All authors are with Division of Mathematical Sciences, School of Physical \& Mathematical Sciences, Nanyang Technological University, Singapore. Emails in sequence: s080001@e.ntu.edu.sg, \{hoonwei,lingsan\}@ntu.edu.sg, lsu1@e.ntu.edu.sg, hxwang@ntu.edu.sg.}}

\maketitle

\begin{abstract}
In Identity-Based Encryption (IBE) systems, key revocation is non-trivial.
This is because a user's identity is itself a public key.
Moreover, the private key corresponding to the identity needs to be obtained from a trusted key authority through an authenticated and secrecy protected channel. So far, there exist only a very small number of revocable IBE (RIBE) schemes that support non-interactive key revocation, in the sense that the user is not required to interact with the key authority or some kind of trusted hardware to renew her private key without changing her public key (or identity).
These schemes are either proven to be only selectively secure or have public parameters which grow linearly in a given security parameter.
In this paper, we present two constructions of non-interactive RIBE that satisfy all the following three attractive properties: (i) proven to be adaptively secure under the Symmetric External
Diffie-Hellman (SXDH) and the Decisional Linear (DLIN) assumptions; (ii) have constant-size public parameters; and (iii) preserve the anonymity of ciphertexts---a property that has not yet been achieved in all the current schemes.
\end{abstract}

\begin{IEEEkeywords}
 Dual System Encryption, Functional Encryption, Identity-Based Encryption, Key Revocation
\end{IEEEkeywords}

\section{Introduction}
%
%
%
%
{\em \IEEEPARstart{I}{}dentity-based encryption} (IBE) allows one's identity to be
directly used as a public key~\cite{Shamir84,BonehF01:ibe,Cocks01}. This
obviates the need for a public key certificate that attests the
binding between the identity and a (seemingly) random key, as in
the case of more conventional certificate-based public-key systems. Thus, IBE
systems have simpler public key management than that of
certificate-based systems. In IBE, however, a private key
(corresponding to an identity) needs to be generated by a trusted
key authority. This and the fact that a user's identity is itself a
public key complicates key renewal or revocation---one cannot simply
change her public key, as this changes her identity as well. While
there has been a great deal of work on IBE in recent years, see for
example~\cite{BonehBG05,Waters05,BonehGH07,GentryPV08,Waters09,CashHKP10,AgrawalBB10a,AgrawalBB10b},
not much work has been devoted to {\em key revocation}.

One direct way to alleviate the key revocation problem in the IBE
setting is to maintain a revocation list by some trusted third party.
A sender checks on the trusted third party and just stops to encrypt messages if the corresponding receiver is revoked.
However, this direct model requires the trusted third party to keep online in order to respond any sender's real time checking query.
To address this problem, one simple solution is to append a validity period to a target identity
during encryption~\cite{BonehF01:ibe}. This results in a public key with
a limited validity period, and hence, restricting the window of
exposure should the corresponding private key is compromised. If the
validity period is sufficiently short, one may not require an
explicit key revocation mechanism since an exposed private key is of
little value to an adversary beyond the specified validity period.
However, one major drawback of this approach is that each user has
to periodically renew her private key. As a consequence, the key
authority's workload increases linearly in the number of non-revoked
users. Further, we must ensure that each transmission of a new
private key between the key authority and a non-revoked user is
performed through some form of authenticated and secure channel.
There exist some improved key revocation techniques in the
literature, for example~\cite{LibertQ03,HanaokaHSI05}.
However, they require interactions either between the user and the key authority, as before,
or between the user and some kind of trusted hardware.
These may not always be practical.

The first non-interactive, revocable IBE (RIBE) scheme that neither
presupposes the existence of trusted hardware nor requires a secure channel between the user and the key authority, is due to Boldyreva et al.~\cite{BoldyrevaGK08}.
Their scheme borrows the concept of fuzzy
IBE (FIBE)~\cite{SahaiW05} in which encryption of a message is associated with two ``attributes'', namely identity of the
receiver and time period. The corresponding decryption key is also
split into two private components, matching the identity and the
time period, respectively. The private component that corresponds to
the identity is essentially similar to a regular private key in IBE
and it is issued to a user by the key authority through a secure
channel. On the other hand, the private key component corresponding
to the time period is regarded as a key update and is published by the key authority to all users. (Here the key update is public information and does not require secrecy protection.) Thus, to revoke a
user, the key authority simply stops distributing the key update for
that user. To reduce the number of key updates to be performed by
the key authority, Boldyreva et al.\ organize and relate users' key
updates in a binary tree~\cite{AielloLO98,NaorN00}. Briefly
speaking, each node of the tree is assigned some key material and
each user is assigned to a leaf node in the tree. Upon registration,
the key authority computes and provides the user with a set of
distinct private keys (corresponding to its identity) based on the
key material for each node in the path from the leaf node
corresponding to that user to the root node. To be able to decrypt a
ciphertext associated with time $\tp$, the user needs just one key
update (corresponding to $\tp$) computed on the key material
associated to any of the nodes on the path from the leaf node of the
user to the root node. Thus, when no user is revoked, the key
authority publishes just the key update computed on the key material
of the root node. When a subset of the users is revoked, the key
authority first finds the minimal set of nodes in the tree which
contains an ancestor (or, the node itself) among all the leaf nodes
corresponding to non-revoked users. The key authority then
distributes the key updates for only this set. This way, every update of
the revocation list only requires the key authority to perform
logarithmic work in the maximal number of users and linear in the
number of revoked users.

\subsection{Previous Non-Interactive RIBE Constructions}
Although an adaptive-ID secure IBE scheme~\cite{Waters05} (which
is resilient even against an adversary that is allowed to adaptively select an identity as the attack target based on the responses to the adversary's queries in a security game) has been in existence for some years, constructing an RIBE scheme with equivalent security guarantee is non-trivial. This is evident from the first RIBE scheme proposed by Boldyreva et al.~\cite{BoldyrevaGK08}. Although it is intriguing that their RIBE scheme was constructed from the FIBE scheme of~\cite{SahaiW05} and made clever use of the binary tree technique, the scheme was only proven in the selective-ID model, which is, unfortunately, a rather weak model. This is because the adversary is required to set the challenge identity and time at the beginning of a security game before receiving the relevant public parameters.
Nevertheless, Libert and Vergnaud~\cite{LibertV09:ribe} eventually proposed an adaptive-ID secure RIBE scheme
using similar key revocation techniques as with~\cite{BoldyrevaGK08}, and thus solved the problem left open by Boldyreva et al.
However, instead of building on FIBE, Libert and Vergnaud adopted a variant~\cite{LibertV09:aibe} of the Waters IBE
scheme~\cite{Waters05}, which is based on partitioning techniques and has a drawback in having public parameters that comprise
$\bigO(\lambda)$ group elements for security parameter $\lambda$.
Consequently, the Libert and Vergnaud RIBE scheme inherits a similar limitation.
Clearly, it is desirable that a scheme has small or constant-size
public parameters, secret keys, and ciphertexts, if it were to be
deployed in real world applications.

\subsection{RIBE from Dual System Encryption}
Moving beyond proving security through the partitioning techniques, Waters proposed the dual system encryption methodology~\cite{Waters09}, which has been a powerful tool to obtain full security for various classes of functional encryption (FE)~\cite{BSW11:fe}, such as IBE~\cite{Waters09,LewkoW10,Lewko12}, inner product encryption (IPE)~\cite{LewkoOSTW10}, and attribute-based encryption (ABE)~\cite{LewkoOSTW10,OkamotoT10}.
Although there already exist some schemes that achieve full security using the dual system encryption technique, (for example, the HIBE scheme of~\cite{LewkoW10} has been proven to be fully secure by applying this technique to the HIBE scheme of~\cite{BonehBG05}), however, these fully secure schemes typically require relatively large parameters and/or constructed only in the composite order bilinear groups. Thus, in general, the dual system encryption methodology does not always provide generic transformation from selective security to adaptive security without suffering from the mentioned limitations.

In our work, we initially tried to apply the dual system encryption technique to the selective-ID RIBE scheme of~\cite{BoldyrevaGK08}, however this results an analogous construction and proof to the ABE scheme of~\cite{OkamotoT10}. Furthermore, as we illustrate below, such an approach does not enjoy constant-size public parameters and keys.

To see this, we specifically consider the binary-tree key update approach~\cite{BoldyrevaGK08,LibertV09:ribe} in the setting similar to key-policy ABE.\footnote{The case for ciphertext-policy ABE setting is similar.}
As before, a ciphertext in the RIBE scheme is associated with two attributes: identity $\idp_i$ and time period $\tp_j$.
The ciphertext can be decrypted by a user if and only if the user possesses both the private key for identity $\idp_i$ and the key update for time $\tp_j$ on some node in the tree.
Since the private keys and key updates associated with a specific node are not given to the users simultaneously, collusion among some (non-revoked) users on some attributes (i.e.\ time
attribute) is possible.
Hence from the view of ABE, all users can be regarded as ``sharing'' the same key (or private component) associated with access structure of the form
\[
(\idp_1\vee\cdots\vee \idp_n)\wedge(\tp_1\vee\cdots\vee \tp_m)
\]
on each node in the tree for some integers $n$ and $m$, but each user is given only some parts of the key for this access structure.
That is, the parts of the key that the user gets correspond to access structure $\idp_i\wedge(\tp_1\vee\cdots\vee \tp_m)$ if this node is in the path from the leaf node associated with $\idp_i$ to the root node; while the key updates corresponding to $(\tp_1\vee\cdots\vee \tp_m)$ are given to all users (not necessarily at the same time).
Clearly, we require that the private keys are collusion-resistant on different nodes.
Moreover, supporting a large universe attribute space is required and can be used to deal with exponential identity spaces in RIBE.

We observe that, however, the adaptively secure ABE schemes of~\cite{OkamotoT10} cannot be used directly for our purpose because the resulting RIBE somewhat unexpectedly has private keys and ciphertexts with sizes that grow linearly in the maximal number of users and the size of time space (even though they are polynomial in the security parameter). It turns out that constructing a fully secure RIBE scheme with constant-size public parameters and keys requires additional work.

\subsection{Our Contributions}
In this paper, we investigate how to instantiate the Waters dual system encryption methodology with revocable IBE schemes.
Particularly, we construct two efficient non-interactive RIBE schemes that are proven to be adaptively secure under the Symmetric External
Diffie-Hellman (SXDH) and the Decisional Linear (DLIN) assumptions, respectively.

Our schemes improve the previous work by achieving adaptive security with {\em constant-size public parameters}.
Moreover, our schemes are {\em anonymous}, namely, preserving the privacy of ciphertext recipients and encryption times.
We note that previous RIBE schemes do not consider the anonymity property, an advantage inherited from using the dual pairing vector spaces (DPVS)~\cite{OkamotoT08,OkamotoT09} to achieve orthogonality and entropy-hiding in prime-order groups.
Our constructions also make use of the key revocation techniques of~\cite{BoldyrevaGK08,LibertV09:ribe},
namely, we employ binary-tree data structure to achieve key update with logarithmic
complexity in the maximal number of users for the key authority.

We give a summary of comparisons between existing and our RIBE schemes in
Table~\ref{tab:comparison-ribe}.
Here, we use $\PP$ to denote public parameters, $\MK$ to denote master key, $\SK$ to denote private key, $\KU$ to denote key update, $\CT$ to denote ciphertext, and \# pairing to denote the number of pairing computation for decryption.
The sizes are in terms of group elements and $\lambda$ denotes the security parameter.

\begin{table}[H]
\centering
\caption{Comparisons between existing and our RIBE
schemes.}\label{tab:comparison-ribe}
\medskip
\begin{tabular}{|l|c|c||c|c|c|c|c|c|c|}
\hline
                       & BGK \cite{BoldyrevaGK08}  & LV \cite{LibertV09:ribe} & \multicolumn{2}{c|}{Ours} \\
\hline
size of $\PP$          &  $5$                      & $\bigO(\lambda)$         & $19$   & $55$  \\
\hline
size of $\MK$          &  $1$                      & $1$                      & $19$   & $55$ \\
\hline
size of $\SK$          &  $2$                      &  $3$                     & $6$    & $9$  \\
\hline
size of $\KU$          &  $2$                      &  $3$                     & $6$   & $9$ \\
\hline
size of $\CT$          &   $4$                     &   $5$                    & $6$   & $9$   \\
\hline
\# pairings            &   $4$                     &  $3$                     & $12$  & $18$ \\
\hline
security               & selective                 & adaptive                 & adaptive & adaptive   \\
\hline
anonymity             & No                        & No                       & Yes & Yes \\
\hline
assumption            & DBDH                      & mDBDH                     &  SXDH  & DLIN  \\
\hline
\end{tabular}
\end{table}

We compare our schemes against Boldyreva et al.'s scheme~\cite{BoldyrevaGK08}, which is under the Decision Bilinear Diffie-Hellman (DBDH) assumption, and Libert and Vergnaud's scheme~\cite{LibertV09:ribe}, which is under the modified DBDH (mDBDH) assumption.
Overall, our schemes are anonymous, adaptively secure, and have constant-size public parameters, at the expense of bigger (but still seems acceptable) sizes in terms of the master key, private key, and key update.

\subsection{Our Approach}
In RIBE, different from the standard security game for IBE, the adversary is allowed to query parts of the challenge identities and time periods.
Thus, to overcome the problem of increasing sizes of public parameters in
the maximal number of users and sizes of the time space as analyzed in the ABE setting, our security proof makes use of two types of nominally semi-functional pairs,
while all the previous works based on the dual system encryption methodology, such as~\cite{LewkoW10,Lewko12,LewkoOSTW10,OkamotoT10}, require only a single type of nominally semi-functional pair.
Moreover, prior to the start of the game, we execute a preliminary game to ``locate'' the positions of
the challenge identities and times. We then transform all the private keys and key updates  associated with the non-challenge identities and times, respectively, into nominally semi-functional (we denote by Type I) one by one. We transform the challenge private keys and key updates (or simply keys) into nominally semi-functional (we denote by Type II) node by node at the last step. Note that the distribution of nominally semi-functional pairs of Type I for challenge identities
and times can be detected by the adversary that they are different
from the distribution of the semi-functional keys and ciphertexts.
Moreover, nominally semi-functional pairs of Type II can be only
generated for the last remaining keys; in other words, all the other
keys must have been already semi-functional.
This is why the preliminary game is needed.
We also introduce some statistical indistinguishability arguments in our proof to show that the distributions of nominally semi-functional pair of both Types I \& II remain
the same as the distributions of semi-functional keys and ciphertexts from the adversary's view.
Finally, we arrive at a
security game that only requires to generate semi-functional keys and ciphertexts while security
can be proved directly.

\section{Preliminaries}
\subsection{Dual Pairing Vector Spaces}
Our constructions are based on dual pairing vector
spaces proposed by Okamoto and Takashima \cite{OkamotoT08,OkamotoT09}.
In this paper, we concentrate on the asymmetric version \cite{OkamotoT10f}.
Particularly, we give a brief description on how to generate random dual orthonormal bases. See \cite{OkamotoT08,OkamotoT09} for a full definition of dual pairing vector
spaces.

\newtheorem{mydef}{Definition}
\newtheorem{mythm}{Theorem}
\newtheorem{mylem}{Lemma}

\begin{mydef}
[Asymmetric bilinear pairing groups] Asymmetric bilinear pairing
groups $(q,G_1,G_2,G_T,g_1,g_2,e)$ are a tuple of a prime $q$,
cyclic (multiplicative) groups $G_1,G_2$ and $G_T$ of order $q$,
$g_1\neq 1\in G_1$, $g_2\neq 1\in G_2$, and a polynomial-time
computable nondegenerate bilinear pairing $e: G_1\times
G_2\rightarrow G_T$ i.e., $e(g_1^s, g_2^t) = e(g_1, g_2)^{st}$ and
$e(g_1, g_2) \neq 1$.
\end{mydef}

In addition to individual elements of $G_1$ or $G_2$,
we will also consider ``vectors'' of group elements. For $\vecv =
(v_1,\ldots, v_n) \in \Z_q^n$ and $g_\beta\in G_\beta$, we write
$g_\beta^{\vecv}$ to denote a $n$-tuple of elements of $G_\beta$ for
$\beta=1,2$:
\[
g_\beta^{\vecv}:=(g_\beta^{v_1},\ldots,g_\beta^{v_n}).
\]
\noindent For any $a\in\Z_q$ and $\vecv,\vecw\in \Z_q^n$, we have:
\[
g_\beta^{a\vecv}:=(g_\beta^{av_1},\ldots,g_\beta^{av_n}), \quad g_\beta^{\vecv+\vecw}:=(g_\beta^{v_1+w_1},\ldots,g_\beta^{v_n+w_n}).
\]
\noindent Then we define
\[
e(g_1^{\vecv},g_2^{\vecw}):=\prod_{i=1}^ne(g_1^{v_i},g_2^{w_i})=e(g_1,g_2)^{\vecv\cdot\vecw}.
\]
\noindent Here, the dot product is taken modulo $q$.

\textit{Dual Pairing Vector Spaces.}
For a fixed (constant) dimension $n$, we choose two random
bases $\Bdual := (\vecb_1,\ldots,\vecb_n)$ and $\Bdual^* :=
(\vecb_1^*,\ldots,\vecb_n^*)$ of $\Z_q^n$, subject to the constraint
that they are ``dual orthonormal'', meaning that
\[
\vecb_i\cdot\vecb_j^*=0\, (\mbox{mod } q)
\]
\noindent whenever $i\neq j$, and
\[
\vecb_i\cdot\vecb_i^*=\psi\, (\mbox{mod } q)
\]
\noindent for all $i$, where $\psi$ is a random element of $\Z_q$.
We denote the above algorithm, which generates the
dual orthonormal bases, as $\Dual(\cdot)$.
Then for generators $g_1\in G_1$ and $g_2\in G_2$, we have
\[
e(g_1^{\vecb_i},g_2^{\vecb_j^*})=1
\]
whenever $i\neq j$, where $1$ here denotes the identity element in $G_T$.

\subsection{Complexity Assumptions}
To define the SXDH assumption, we first define DDH problems in $G_1$ and $G_2$.

\begin{mydef}
[DDH1: Decisional Diffie-Hellman Assumption in $G_1$]\label{def:ddh1} Given a group generator $\mathcal{G}$, we define the
following distribution:
\begin{align*}
&\mathbb{G}:=(q,G_1,G_2,G_T,g_1,g_2,e)\getsr\mathcal{G},\\
&a,b,c \getsr\Z_q,\\
&D:=(\mathbb{G};g_1,g_2,g_1^a,g_1^b).
\end{align*}
\noindent We assume that for any PPT algorithm $\AdvA$ (with output in $\{0, 1\}$),
\[
\Adv_{\AdvA}^{\DDHop}(\lambda):=\left| \Pr[\AdvA(D,g_1^{ab})-\Pr[\AdvA(D,g_1^{ab+c})]\right|
\]
\noindent is negligible in the security parameter $\lambda$.
\end{mydef}

\noindent The dual of the Decisional Diffie-Hellman assumption in $G_1$ is Decisional Diffie-Hellman assumption in $G_2$ (denoted as DDH2), which is identical to Definitions~\ref{def:ddh1} with the roles of $G_1$ and $G_2$ reversed.
We say that:

\begin{mydef}
The Symmetric External Diffie-Hellman assumption holds if DDH problems are intractable in both
$G_1$ and $G_2$.
\end{mydef}

The following SXDH-based Subspace assumptions is from \cite{ChenLLWW12:ibe-sxdh}, which we will use in our security proof.

\begin{mydef}
[DS1: Decisional Subspace Assumption in $G_1$]\label{def:ds1}
Given a group generator $\mathcal{G}(\cdot)$, define the following distribution:
\begin{align*}
&\quad\quad \mathbb{G}:=(q,G_1,G_2,G_T,g_1,g_2,e)\getsr\mathcal{G}(1^{\lambda}),\\
&\quad\quad (\Bdual,\Bdual^*)\getsr \Dual(\Z_q^\conNp),\\
&\quad\quad \tau_1,\tau_2,\mu_1,\mu_2\getsr\Z_q,\\
&\quad\quad U_1:=g_2^{\mu_1\vecb_1^*+\mu_2\vecb_{\conKp+1}^*},\ldots,U_\conKp:=g_2^{\mu_1\vecb_\conKp^*+\mu_2\vecb_{2\conKp}^*},\\
&\quad\quad V_1:=g_1^{\tau_1\vecb_1},\ldots,
V_\conKp:=g_1^{\tau_1\vecb_\conKp},\\
&\quad\quad W_1:=g_1^{\tau_1\vecb_1+\tau_2\vecb_{\conKp+1}},\ldots,
W_\conKp:=g_1^{\tau_1\vecb_\conKp+\tau_2\vecb_{2\conKp}},\\
&\quad\quad D:=(\mathbb{G};g_2^{\vecb_1^*},\ldots,g_2^{\vecb_\conKp^*},g_2^{\vecb_{2\conKp+1}^*},\ldots,
g_2^{\vecb_\conNp^*},\\
&\quad\quad\quad\quad\quad\quad g_1^{\vecb_1},\ldots,g_1^{\vecb_\conNp},U_1,\ldots,U_\conKp,\mu_2),
\end{align*}
\noindent where $\conKp, \conNp$ are fixed positive integers that satisfy $2\conKp\leq \conNp$. We assume that for any PPT algorithm $\AdvA$ (with output in $\{0,1\}$),
\begin{equation*}
\Adv_{\AdvA}^{\DSop}(\lambda):= |\Pr[\AdvA(D,V_1,\ldots,V_\conKp)=1] - \Pr[\AdvA(D,W_1,\ldots,W_\conKp)=1]|
\end{equation*}

\noindent is negligible in the security parameter $\lambda$.
\end{mydef}

\begin{mylem}\label{lem:DS1}
If the DDH assumption in $G_1$ holds, then the Subspace
assumption in $G_1$ stated in Definition \ref{def:ds1} also holds. More precisely, for any adversary $\AdvA$ against the Subspace
assumption in $G_1$, there exist probabilistic algorithms $\AdvB$ whose running times
are essentially the same as that of $\AdvA$, such that
\[
\Adv_{\AdvA}^{\DSop}(\lambda)\leq \Adv_{\AdvB}^{\DDHop}(\lambda).
\]
\end{mylem}

\noindent The dual of the Subspace assumption in $G_1$ is Subspace assumption in $G_2$ (denoted as DS2), which is identical to Definitions~\ref{def:ds1} with the roles of $G_1$ and $G_2$ reversed.
Similarly, the Subspace assumption holds in $G_2$ if the DDH assumption in $G_2$ holds.

We define the DLIN problem in symmetric bilinear pairing groups (namely $G_1=G_2$). The DLIN-based Subspace assumptions
could be found in \cite{Lewko12,OkamotoT10}.

\begin{mydef}
[DLIN: Decisional Linear Assumption] Given a group generator $\mathcal{G}$, we define the
following distribution:
\begin{align*}
&\mathbb{G}:=(q,G,G_T,g,e)\getsr\mathcal{G},\\
&a_1,a_2,b_1,b_2,c \getsr\Z_q,\\
&D:=(\mathbb{G};g,g^{a_1},g^{a_2},g^{a_1b_1},g^{a_2b_2}).
\end{align*}
\noindent We assume that for any PPT algorithm $\AdvA$ (with output in $\{0, 1\}$),
\[
\Adv_{\AdvA}^{\mathsf{DLIN}}(\lambda):=\left| \Pr[\AdvA(D,g^{b_1+b_2})-\Pr[\AdvA(D,g_1^{b_1+b_2+c})]\right|
\]
\noindent is negligible in the security parameter $\lambda$.
\end{mydef}

\section{Revocable IBE} \label{definitions}

We first recall the definition of RIBE and its security
from~\cite{BoldyrevaGK08} and then define an appropriate security
model for our constructions.
\begin{mydef}
An Identity-Based Encryption with efficient revocation or simply
Revocable IBE (RIBE) scheme has seven PPT algorithms $\Setup$, $\PriKeyGen$, $\KeyUpd$, $\DecKeyGen$, $\Enc$, $\Dec$, and $\KeyRev$ with associated message space $\Msgsp$, identity space $\Idsp$ and time space $\Tmsp$.
We assume that the size of $\Tmsp$
is polynomial in the security parameter. Each algorithm is run by either one of three types of
parties---key authority, sender or receiver. Key authority maintains a revocation list $\RLp$ and state $\STp$.
In what follows,
an algorithm is called stateful if it updates $\RLp$ or $\STp$. We treat time as discrete as opposed to continuous.

\begin{IEEEitemize}
\item $\Setup(1^\lambda,N_{max})$  takes as input a security parameter $\lambda$ and a maximal
number of users $N_{max}$. It outputs public parameters $\PP$, a master key $\MK$, a revocation list $\RLp$ (initially empty), and a state $\STp$.
(This is run by the key authority.)

\item $\PriKeyGen(\PP,\MK,\idp,\STp)$  takes as input the public parameters $\PP$, the master key $\MK$, an identity $\idp\in \Idsp$, and the state $\STp$. It outputs a private key $\SK_{\idp}$ and an updated state $\STp$. (This is stateful and run by the key authority.)

\item $\KeyUpd(\PP,\MK,\tp,\RLp,\STp)$ takes as input the public parameters
$\PP$, the master key $\MK$, a key update time $\tp\in\Tmsp$, the revocation list $\RLp$, and the state $\STp$. It outputs a key update $\KU_\tp$. (This is run by the key authority.)

\item $\DecKeyGen(\SK_{\idp},\KU_\tp)$ takes as input a private key $\SK_{\idp}$
and key update $\KU_\tp$. It outputs a decryption key $\DK_{\idp,\tp}$ or a special symbol $\bot$ indicating
that $\idp$ was revoked. (This is deterministic and run by the receiver.)

\item $\Enc(\PP,\idp,\tp,\msgp)$ takes as input the public parameters $\PP$, an identity $\idp\in \Idsp$, an encryption time $\tp\in \Tmsp$, and a message $\msgp \in \Msgsp$. It outputs a ciphertext $\CT_{\idp,\tp}$. (This is run by the sender.)

\item $\Dec(\PP,\DK_{\idp,\tp},\CT_{\idp,\tp})$ takes as input the public parameters $\PP$, a decryption key $\DK_{\idp,\tp}$, and a ciphertext $\CT_{\idp,\tp}$. It outputs a message $\msgp\in \Msgsp$. (This is deterministic and run by the receiver.)

\item $\KeyRev(\idp,\tp,\RLp,\STp)$ takes as input an identity to be revoked
$\idp\in \Idsp$, a revocation time $\tp\in \Tmsp$, the revocation list $\RLp$, and the state $\STp$. It outputs an updated revocation list $\RLp$. (This is stateful and run by the key authority.)
\end{IEEEitemize}
\end{mydef}

\noindent The consistency condition requires that for all $\lambda\in \mathbb{N}$ and polynomials (in $\lambda$) $N_{max}$, all $\PP$ and $\MK$ output
by setup algorithm \textsf{\textup{Setup}}, all $\msgp \in \Msgsp, \idp\in \Idsp, \tp\in \Tmsp$ and all possible valid states $\STp$ and revocation lists $\RLp$, if identity $\idp$ was not revoked before or, at time $\tp$ then the following experiment returns $1$ except for a negligible probability:
\begin{align*}
&(\SK_{\idp}, \STp) \getsr \PriKeyGen(\PP,\MK,\idp,\STp);\\
&\KU_\tp \getsr \KeyUpd(\PP,\MK,\tp,\RLp,\STp)\\
&\DK_{\idp,\tp} \leftarrow \DecKeyGen(\SK_{\idp},\KU_\tp);\\
&\CT_{\idp,\tp} \getsr \Enc(\PP,\idp,\tp,\msgp)\\
&\mbox{If } \Dec(\PP,\DK_{\idp,\tp},\CT_{\idp,\tp})=\msgp \mbox{ then return } 1 \mbox{ else return } 0.
\end{align*}

Boldyreva et al.\ formalized and defined the selective-ID security for RIBE. Their definition captures not only
the standard notion of selective-ID security but also takes into account key revocation. The following definition extends the security
property expressed in~\cite{BoldyrevaGK08} to the adaptive-ID and anonymous setting.

\begin{IEEEitemize}
\item $\SimSetup$: It is run to generate public
    parameters $\PP$, a master key $\MK$, a revocation list $\RLp$ (initially empty), and a state $\STp$. Then $\PP$ is given to $\AdvA$.

\item $\SimQuery$: $\AdvA$ may adaptively make a
polynomial number of queries of the following oracles (the oracles share state):

\begin{IEEEitemize}
\item The private key generation oracle $\PriKeyGen(\cdot)$ takes as input an identity $\idp$ and runs $\PriKeyGen(\PP,\MK,\idp,\STp)$ to return a private key $\SK_{\idp}$.
\item The key update generation oracle $\KeyUpd(\cdot)$ takes as input time $\tp$ and runs $\KeyUpd(\PP,\MK,\tp,\RLp,\STp)$ to
return a key update $\KU_\tp$.
\item The revocation oracle $\KeyRev(\cdot,\cdot)$ takes as input an identity $\idp$ and time $\tp$ and runs $\KeyRev(\idp,\tp,\RLp,\STp)$ to update $\RLp$.
\end{IEEEitemize}

\item $\SimChallenge$: $\AdvA$ outputs the two challenge pair $(\idp^*_{(0)},\tp^*_{(0)},\msgp^*_{(0)}),(\idp^*_{(1)},\tp^*_{(1)},\msgp^*_{(1)})$ $\in\Idsp\times \Tmsp\times\Msgsp$.
A random bit $\beta$ is chosen. $\AdvA$ is given
$\Enc(\PP,\idp^*_{(\beta)},\tp^*_{(\beta)},\msgp^*_{(\beta)})$.

\item $\SimGuess$: The adversary may continue to make queries of the oracles as in $\SimQuery$ phase and outputs a bit $\beta^{\prime}$, and succeeds if $\beta^{\prime} = \beta$.
\end{IEEEitemize}

\noindent The following restrictions must always hold:

\begin{IEEEenumerate}
\item $\KeyUpd(\cdot)$ and $\KeyRev(\cdot,\cdot)$ can be queried on time which is greater than or equal to the time of all previous queries, i.e., the adversary is allowed to query only in non-decreasing order of time. Also, the oracle $\KeyRev(\cdot,\cdot)$ cannot be queried at time $\tp$ if $\KeyUpd(\cdot)$ was queried on $\tp$.
\item For $\beta=0,1$, if $\PriKeyGen(\cdot)$ was queried on identity $\idp_{(\beta)}$ then $\KeyRev(\cdot,\cdot)$ must be queried on $(\idp^*_{(\beta)},\tp)$ for some $\tp\leq \tp^*_{(\beta)}$, i.e., identity $\idp^*_{(\beta)}$ must be in $\RLp$ when $\KeyUpd(\cdot)$ is queried at time $\tp^*_{(\beta)}$.
\end{IEEEenumerate}

\noindent For $\beta=0,1$ let $W_{\beta}$ be the event that the adversary outputs $1$ in Experiment $\beta$ and define
\[
\Adv_{\AdvA}^{\RIBEp}(\lambda):=|\Pr[W_0]-\Pr[W_1]|.
\]

\begin{mydef}\label{def:secrutiy}
An RIBE scheme is adaptive-ID secure and anonymous if for all PPT adversaries $\AdvA$ the function $\Adv_{\AdvA}^{\RIBEp}(\lambda)$ is
negligible.
\end{mydef}

\Remark
The security notion of {\em non-anonymous} RIBE is defined as above with
restriction that $\idp^*_{(0)}=\idp^*_{(1)}$ and
$\tp^*_{(0)}=\tp^*_{(1)}$. On the other hand, if the adversary $\AdvA$
outputs $(\idp^*_{(0)},\idp^*_{(0)})$ and
$(\idp^*_{(1)},\tp^*_{(1)})$ before the $\Setup$ phase, it is
{\em selective-ID} security.

\section{Construction from SXDH}\label{sxdh_construction}

In this section, we present our first construction of RIBE and its proof of security under the SXDH assumption.

\subsection{The Binary-tree Data Structure}
Key revocation in our scheme relies on binary-tree data structure, as
with~\cite{AielloLO98,NaorN00,BoldyrevaGK08,LibertV09:ribe}. We
denote the binary-tree by $\BT$ and its root node by $\Root$. If
$\nu$ is a leaf node then $\Path(\nu)$ denotes the set of nodes on
the path from $\nu$ to $\Root$ (both $\nu$ and $\Root$ inclusive).
If $\theta$ is a non-leaf node then $\theta_{\ell}$, $\theta_r$
denote the left and right child of $\theta$, respectively. We assume
that all nodes in the tree are uniquely encoded as strings, and the
tree is defined by all of its node descriptions.

Each user is assigned to a leaf node $\nu$. Upon registration, the
key authority provides the user with a set of distinct private keys
for each node in $\Path(\nu)$.
At time $\tp$, the key authority uses an algorithm called $\KUNodes$ to determine the minimal set $\mathsf{Y}$ of nodes in $\BT$ such that none of the nodes in $\RLp$
with corresponding time $\leq \tp$ (users revoked on or before
$\tp$) have any ancestor (or, themselves) in the set $\mathsf{Y}$,
and all other leaf nodes (corresponding to non-revoked users) have
exactly one ancestor (or, themselves) in the set. The $\KUNodes$ algorithm takes as input a binary tree $\BT$, a revocation list $\RLp$ and a time $\tp$, and can be formally
specified as follows:
\begin{align*}
&\KUNodes(\BT,\RLp,\tp)\\
&\quad \mathsf{X},\mathsf{Y}\leftarrow \emptyset\\
&\quad \forall(\nu_i,\tp_i)\in\RLp\\
&\quad\quad \mbox{if } \tp_i\leq \tp \mbox{ then add } \Path(\nu_i) \mbox{ to } \mathsf{X}\\
&\quad \forall \theta \in \mathsf{X}\\
&\quad\quad \mbox{if } \theta_{\ell}\not \in \mathsf{X} \mbox{ then add } \theta_{\ell} \mbox{ to } \mathsf{Y}\\
&\quad\quad \mbox{if } \theta_{r}\not \in \mathsf{X} \mbox{ then add } \theta_{r} \mbox{ to } \mathsf{Y}\\
&\quad \mbox{If } \mathsf{Y}=\emptyset \mbox{ then add } \Root\mbox{ to } \mathsf{Y}\\
&\quad \mbox{Return } \mathsf{Y}
\end{align*}
\noindent The $\KUNodes$ algorithm marks all the ancestors of
revoked nodes as revoked and outputs all the non-revoked children of
revoked nodes.
The key authority then publishes a key update for all nodes of
$\mathsf{Y}$.
A user assigned to leaf $\nu$ is then able to form an effective
decryption key for time $\tp$ if the set $\mathsf{Y}$ contains a
node in $\Path(\nu)$. By doing so, every update of the revocation
list $\RLp$ only requires the key authority to perform logarithmic
work in the maximal number of users and linear in the number of
revoked users.

\subsection{Our Scheme}

We now specify our RIBE scheme. We sometimes provide some intuition or remark at the end of an algorithm and this is marked by the symbol ``//''.

\begin{IEEEitemize}
\item $\Setup(\lambda,N_{max})$ On input a security parameter $\lambda$, and a maximal number $N_{max}$ of users, and generate a
bilinear pairing $\mathbb{G}:=(q,G_1,G_2,G_T,g_1,g_2,e)$ for sufficiently large prime order $q$. Next perform the following steps:
\begin{IEEEenumerate}
\item Let $\RLp$ be an empty set and $\BT$ be a binary-tree with at least $N_{max}$ leaf nodes, set $\STp=\BT$.

\item Sample random dual orthonormal bases, $(\Ddual,\Ddual^\ast)\getsr \Dual(\Z_q^\dldim)$. Let $\vecd_1,\ldots,\vecd_{\dldim}$ denote the elements of $\Ddual$ and $\vecd_1^\ast,\ldots,\vecd_{\dldim}^\ast$ denote the elements of $\Ddual^\ast$. It also picks $\alpha\getsr \Z_q$ and computes $g_T^\alpha:=e(g_1,g_2)^{\alpha\vecd_1\cdot \vecd_1^\ast}$

\item Output $\RLp$, $\STp$, the public parameters
\begin{align*}
\PP:=\left\{\mathbb{G};g_T^\alpha,g_1^{\vecd_1},g_1^{\vecd_2},g_1^{\vecd_3}\right\},
\end{align*}
and the master key $\MK$
\[
\MK:=\left\{\alpha,g_2^{\vecd_1^\ast},g_2^{\vecd_2^\ast},g_2^{\vecd_3^\ast}\right\}.
\]
\end{IEEEenumerate}

\item $\PriKeyGen(\PP,\MK,\idp,\RLp,\STp)$  On input the public parameters $\PP$, the master key $\MK$, an identity $\idp$, the revocation list $\RLp$, and the state $\STp$, it picks an unassigned leaf node $v$ from $\BT$ and stores $\idp$ in that node. It then performs the following steps:
\begin{IEEEenumerate}
    \item For any $\theta\in \Path(v)$, if $\alpha_{\theta,1},\alpha_{\theta,2}$ are undefined, then pick $\alpha_{\theta,1}\getsr \Z_q$, set $\alpha_{\theta,2}=\alpha - \alpha_{\theta,1}$, and store them in node $\theta$\footnote{To avoid having to store $\alpha_{\theta,1},\alpha_{\theta,2}$ for each node, the authority can derive them from a pseudo-random function of using a shorter seed and re-compute them when necessary.}. Pick $r_{\theta,1}\getsr \Z_q$ and compute
        \[
        \sKp_{\idp,\theta}:=g_2^{ ( \alpha_{\theta,1} + r_{\theta,1} \idp ) \vecd_1^\ast - r_{\theta,1} \vecd_2^\ast }.
        \]

    \item Output $\SK_\idp:=\{(\theta,\sKp_{\idp,\theta})\}_{\theta\in \Path(v)}$, $\STp$.
\end{IEEEenumerate}

//The algorithm computes the $\idp$-component of the decryption key for all the nodes on the path from the leaf node (corresponding to $\idp$) to $\Root$.

\item $\KeyUpd(\PP,\MK,\tp,\RLp,\STp)$ On input the public parameters $\PP$, the master key $\MK$, a time $\tp$, the revocation list $\RLp$, and the state $\STp$, it performs the following steps:
\begin{IEEEenumerate}
\item $\forall \theta\in\KUNodes(\BT, \RLp, \tp)$, if $\alpha_{\theta,1},\alpha_{\theta,2}$ are undefined, then pick $\alpha_{\theta,1}\getsr \Z_q$, set $\alpha_{\theta,2}=\alpha - \alpha_{\theta,1}$, and store them in node $\theta$. Pick $r_{\theta,2}\getsr \Z_q$ and compute
        \[
        \sKp_{\tp,\theta}:=g_2^{ ( \alpha_{\theta,2} + r_{\theta,2} \tp ) \vecd_1^\ast - r_{\theta,2} \vecd_3^\ast }.
        \]
\item
Output $\KU_\tp:=\{(\theta,\sKp_{\tp,\theta})\}_{\theta\in\KUNodes(\BT, \RLp, \tp)}$.
\end{IEEEenumerate}

//The algorithm first finds a minimal set of nodes which contains an ancestor (or, the node
itself) of all the non-revoked nodes. Then it computes the $\tp$-component of the decryption
key for all the nodes in that set.

\item $\DecKeyGen(\SK_{\idp},\KU_\tp)$ On input a private secret key $\SK_{\idp}:=\{(i,\sKp_{\idp,i})\}_{i \in \mathsf{I}}$, $\KU_\tp:=\{(j,\sKp_{\tp,j})\}_{j\in \mathsf{J}}$ for some set of nodes $\mathsf{I},\mathsf{J}$, it runs the following steps:
\begin{IEEEenumerate}
\item $\forall (i,\sKp_{\idp,i})\in \SK_{\idp},(j,\sKp_{\tp,j})\in\KU_\tp$, if $\exists (i, j)$  s.t.\ $i=j$ then $\DK_{\idp,\tp}\leftarrow (\sKp_{\idp,i},\sKp_{\tp,j})$;
else (if $\SK_{\idp}$ and $\KU_\tp$ do not have any node in common) $\DK_{\idp,\tp}\leftarrow \bot$.
\item Output $\DK_{\idp,\tp}$.
\end{IEEEenumerate}


\item $\Enc(\PP,\idp,\tp,\msgp)$ On input the public parameters $\PP$, an identity $\idp$, a time $\tp\in \Z_q^n$, and a message $\msgp$, pick $z\getsr\Z_q$ and forms the ciphertext as
\begin{align*}
\CT_{\idp,\tp}:=\left\{
\sCp:=\msgp\cdot (g_T^{\alpha})^z, \quad \sCp_0:=g_1^{ z ( \vecd_1 + \idp \vecd_2+ \tp \vecd_3 ) }
\right\}.
\end{align*}

\item $\Dec(\PP,\DK_{\idp,\tp},\CT_{\idp,\tp})$  On input the public parameters $\PP$, a decryption key $\DK_{\idp,\tp}:= (\sKp_{\idp,\theta},\sKp_{\tp,\theta})$, and a ciphertext $\CT_{\idp,\tp}:= (\sCp, \sCp_0)$, it computes the message as
        \[
            \msgp:=\sCp/\left(e(\sCp_0,\sKp_{\idp,\theta})\cdot e(\sCp_0,\sKp_{\tp,\theta})\right).
        \]

\item $\KeyRev(\idp,\tp,\RLp,\STp)$  On input an identity $\idp$, a time $\tp$, the revocation list $\RLp$, and the state $\STp$, the algorithm adds $(\idp, \tp)$ to $\RLp$ for all nodes $\nu$ associated with identity $\idp$ and returns $\RLp$.
\end{IEEEitemize}

This ends the description of our scheme.

\noindent \textit{Correctness:} Observe that
\begin{align*}
&e(\sCp_0,\sKp_{\idp,\theta})\\
=\quad&e(g_1^{ z ( \vecd_1 + \idp \vecd_2 + \tp \vecd_3 ) },g_2^{( \alpha_{\theta,1} + r_{\theta,1} \idp)\vecd_1^\ast - r_{\theta,1} \vecd_2^\ast  } )\\
=\quad&e(g_1,g_2)^{\alpha_{\theta,1} z\vecd_1\cdot \vecd_1^\ast} \cdot e(g_1,g_2)^{z r_{\theta,1} \idp \vecd_1\cdot\vecd_1^\ast- z r_{\theta,1} \idp \vecd_2\cdot\vecd_2^\ast} \\
=\quad&e(g_1,g_2)^{\alpha_{\theta,1} z\vecd_1\cdot \vecd_1^\ast}.
\end{align*}
Similarly,
$e(\sCp_0,\sKp_{\tp,\theta})=e(g_1,g_2)^{\alpha_{\theta,2}
z\vecd_1\cdot \vecd_1^\ast}$. The message is recovered as:
\begin{align*}
&\sCp/e(\sCp_0,\sKp_{\idp,\theta})\cdot e(\sCp_0,\sKp_{\tp,\theta}) \\
=\quad &\msgp\cdot (e(g_1,g_2)^{\alpha\vecd_1\cdot \vecd_1^\ast})^z/e(g_1,g_2)^{\alpha z\vecd_1\cdot \vecd_1^\ast}\\
=\quad &\msgp.
\end{align*}

\subsection{Proof of Security}
\textit{Statistical Indistinguishability Lemmas}:
We require the following two lemmas, which are derived from~\cite{OkamotoT10f}, for our security proofs.

\begin{mylem}\label{lem:mat-indist}
For $p\in\Z_q$, let
\[
C_p:=\left\{(\vecx,\vecv)| \vecx\cdot\vecv=p, \veczero\neq \vecx,\veczero\neq \vecv\in \Z_q^n \right\}.
\]
For all $(\vecx,\vecv)\in C_{p}$, for all $(\vecz,\vecw)\in C_{p}$, and $\matA\getsr
\Z_q^{n\times n}$ ($\matA$ is invertible with overwhelming probability),
\[
\Pr[ \vecx \matA^\tran=\vecz ~\wedge~ \vecv \matA^{-1} =\vecw ]=\frac{1}{\#C_p}.
\]
\end{mylem}

\begin{mylem}\label{lem:mat-indist1}
For $p_1,p_2\in\Z_q$, let
\[
C_{p_1,p_2}:=\left\{(\vecx,\vecv_1,\vecv_2)\left|
\begin{array}{l}
 \vecx \neq 0, \vecx\cdot\vecv_1=p_1,\vecx\cdot\vecv_2=p_2  
\end{array}\right. \right\}
\]
where $~\vecx,\vecv_1,\vecv_2\in \Z_q^n, \{\vecv_1,\vecv_2\}$ are linearly independent over $\Z_q$. For all $(\vecx,\vecv_1,\vecv_2)\in C_{p_1,p_2}$, for all $(\vecz,\vecw_1,\vecw_2)\in C_{p_1,p_2}$, and $\matA\getsr
\Z_q^{n\times n}$ ($\matA$ is invertible with overwhelming probability),
\[
\Pr[ \vecx \matA^\tran=\vecz ~\wedge~ \vecv_1 \matA^{-1} =\vecw_1 ~\wedge~ \vecv_2 \matA^{-1} =\vecw_2]=\frac{1}{\#C_{p_1,p_2}}.
\]
\end{mylem}

The following theorem shows that our RIBE scheme is indeed adaptively secure and anonymous.

\begin{mythm}\label{thm:security_ribe}
The RIBE scheme is adaptively secure and anonymous under the SXDH
assumption. More precisely, for any adversary $\AdvA$ against the
RIBE scheme, there exist probabilistic algorithms
\begin{align*}
&\AdvB_0,\\
&\{\AdvB_{\kappa_1,\kappa_2}\}_{\kappa_1=1,\ldots,\qnop,\kappa_2=1,\ldots,\lceil
\log{N_{max}}\rceil},\\
&\{\AdvB_{\kappa_1,\kappa_2}\}_{\kappa_1=\qnop+1,\ldots,\qnop+\qntp+1,\kappa_2=1,\ldots,N_{max}},\\
&\{\AdvB_{\qnop+\qntp+1,\kappa_2}\}_{\kappa_2=1,\ldots,4N_{max}}
\end{align*}
whose running times are essentially the same as that of $\AdvA$,
such that
\begin{align*}
&\Adv_{\AdvA}^{\RIBEp}(\lambda) \leq (\qnop \qntp)^2\cdot \biggl( \Adv_{\AdvB_0}^{\DDHop}(\lambda) + \sum_{\kappa_1=1}^{\qnop}\sum_{\kappa_2=1}^{\lceil \log{N_{max}}\rceil}\Adv_{\AdvB_{\kappa_1,\kappa_2}}^{\DDHtp}(\lambda) + \sum_{\kappa_1=\qnop+1}^{\qntp}\sum_{\kappa_2=1}^{N_{max}}\Adv_{\AdvB_{\kappa_1,\kappa_2}}^{\DDHtp}(\lambda)\biggr.\\
& \biggl.\quad\quad\quad\quad\quad + \sum_{\kappa_2=1}^{4N_{max}}\Adv_{\AdvB_{\qnop+\qntp+1,\kappa_2}}^{\DDHtp}(\lambda) + \frac{6(\qnop \lceil \log{N_{max}}\rceil +\qntp N_{max} ) + 32 N_{max} +6}{q} \biggr)
\end{align*}
where $\qnop,\qntp\geq 4$ are the maximum number of $\AdvA$'s private key
and key update queries respectively.
\end{mythm}

\begin{IEEEproof}
We adopt the dual system encryption methodology by Waters \cite{Waters09} to prove the security of our RIBE scheme.
We use the concepts of {\em semi-functional ciphertexts} and {\em semi-functional keys} in our proof and provide algorithms that generate them.
Particularly, we define two types of semi-functional keys: {\em semi-functional private keys} (for identity) and {\em semi-functional key updates} (for time).
We note that the algorithms (we specify below) are only provided for definitional purposes, and are not part of
the RIBE system. In particular, they do not need to be efficiently computable from the public
parameters and the master key.

\textbf{PriKeyGenSF} The algorithm picks $r_{\theta,1},\nu_{\theta,4,1},\nu_{\theta,5,1},\nu_{\theta,6,1}$ randomly from $\Z_q$ and forms a semi-functional private key for node $\theta$ as
\begin{align}\label{equ:semi-key-ribe-1}
\sKp_{\idp,\theta}^{(\SFp)}:=g_2^{ ( \alpha_{\theta,1} + r_{\theta,1} \idp ) \vecd_1^\ast - r_{\theta,1} \vecd_2^\ast + [\nu_{\theta,4,1}\vecd_4^\ast + \nu_{\theta,5,1}\vecd_5^\ast + \gamma_{\theta,6,1}\vecd_6^\ast] }.
\end{align}

\textbf{KeyUpdSF} The algorithm picks $r_{\theta,2},\nu_{\theta,4,2},\nu_{\theta,5,2},\nu_{\theta,6,2}$ randomly from $\Z_q$ and forms a semi-functional updated key for node $\theta$ as
\begin{align}\label{equ:semi-key-ribe-2}
        \sKp_{\tp,\theta}^{(\SFp)}:=g_2^{ ( \alpha_{\theta,2} + r_{\theta,2} \tp ) \vecd_1^\ast - r_{\theta,2} \vecd_3^\ast  + [\nu_{\theta,4,2}\vecd_4^\ast + \nu_{\theta,5,2}\vecd_5^\ast +
        \nu_{\theta,6,2}\vecd_6^\ast]}.
\end{align}

\textbf{EncryptSF} The algorithm picks $z,\chi_4,\chi_5,\chi_6$ randomly from $\Z_q$ and forms a semi-functional ciphertext as
\begin{align}\label{equ:semi-ct-ribe}
\CT_{\idp,\tp}^{(\SFp)}&:=\left\{\sCp:=\msgp\cdot (g_T^{\alpha})^z, \sCp_0:=g_1^{z( \vecd_1 + \idp \vecd_2 +\tp\vecd_3 ) + (\chi_4\vecd_4 + \chi_5\vecd_5 + \chi_6\vecd_6)}\right\}.
\end{align}

We call a private key or key update semi-functional if all its parts are semi-functional, which are denoted as
\begin{align*}
\SK_\idp^{(\SFp)}&:=\{(\theta,\sKp^{(\SFp)}_{\idp,\theta})\}_{\theta\in \Path(v)}\\
\KU_\tp^{(\SFp)}&:=\{(\theta,\sKp^{(\SFp)}_{\tp,\theta})\}_{\theta\in\KUNodes(\BT, \RLp, \tp)}.
\end{align*}

We observe that a normal ciphertext $\CT_{\idp,\tp}$ can be
decrypted by a semi-functional key pair
$(\sKp_{\idp,\theta}^{(\SFp)},\sKp_{\tp,\theta}^{(\SFp)})$ on some
node $\theta$, because $\vecd_4^\ast,\vecd_5^\ast,\vecd_6^\ast$ are
orthogonal to all of the vectors in exponent of $\sCp_0$, and hence
have no effect on decryption. Similarly, decryption of a
semi-functional ciphertext $\CT_{\idp,\tp}^{(\SFp)}$ by a normal key
pair $(\sKp_{\idp,\theta},\sKp_{\tp,\theta})$ on some node $\theta$
will also succeed because $\vecd_4,\vecd_5,\vecd_6$ are orthogonal
to all of the vectors in the exponent of the key. When both the
ciphertext and key pair on some node are semi-functional, the result
of $e(\sCp_0^{(\SFp)},\sKp_{\idp,\theta}^{(\SFp)})\cdot
e(\sCp_0^{(\SFp)},\sKp_{\tp,\theta}^{(\SFp)})$ will have an
additional term, namely
\[
e(g_1,g_2)^{\sum_{i=4}^6(\nu_{\theta,i,1}+\nu_{\theta,i,2})\chi_i\vecd_i^\ast\cdot \vecd_i}=e(g_1,g_2)^{\sum_{i=4}^6(\nu_{\theta,i,1}+\nu_{\theta,i,2})\chi_i\psi}.
\]
\indent \indent Decryption will then fail unless $\sum_{i=4}^6(\nu_{\theta,i,1}+\nu_{\theta,i,2})\chi_i\psi\equiv 0 ~\mdp q$. If this modular equation holds, we say
that the private key, key update and ciphertext pair is {\em nominally semi-functional}.
In our security proof, there are two types of nominally semi-functional pairs:

\noindent \textbf{Nominally semi-functional pair of Type I}
\begin{align*}
&\sKp_{\idp,\theta}^{(\SFp)}:=g_2^{ ( \alpha_{\theta,1} + r_{\theta,1} \idp ) \vecd_1^\ast - r_{\theta,1} \vecd_2^\ast + [\nu_{\theta,4,1}\idp\vecd_4^\ast - \nu_{\theta,4,1}\vecd_5^\ast] },\\
&\sKp_{\tp,\theta}^{(\SFp)}:=g_2^{ ( \alpha_{\theta,2} + r_{\theta,2} \tp ) \vecd_1^\ast - r_{\theta,2} \vecd_3^\ast  + [\nu_{\theta,4,2}\tp\vecd_4^\ast - \nu_{\theta,4,2}\vecd_6^\ast] },\\
&\CT_{\idp,\tp}^{(\SFp)}:=\left\{\sCp:=\msgp\cdot (g_T^{\alpha})^z, \sCp_0:=g_1^{z( \vecd_1 + \idp \vecd_2 +\tp\vecd_3 ) + [\chi_4( \vecd_4 + \idp \vecd_5 +\tp \vecd_6 )] }\right\},
\end{align*}
where $r_{\theta,1},\nu_{\theta,4,1},r_{\theta,2},\nu_{\theta,4,2},z,\chi_4\getsr \Z_q$.

\noindent \textbf{Nominally semi-functional pair of Type II}
\begin{align*}
&\sKp_{\idp,\theta}^{(\SFp)}:=g_2^{ ( \alpha_{\theta,1} + r_{\theta,1} \idp ) \vecd_1^\ast - r_{\theta,1} \vecd_2^\ast + [(\alpha_{\theta} +\nu_{\theta,4,1}\idp)\vecd_4^\ast - \nu_{\theta,4,1}\vecd_5^\ast] },\\
&\sKp_{\tp,\theta}^{(\SFp)}:=g_2^{ ( \alpha_{\theta,2} + r_{\theta,2} \tp ) \vecd_1^\ast - r_{\theta,2} \vecd_3^\ast  + [(-\alpha_{\theta}+\nu_{\theta,4,2}\tp)\vecd_4^\ast - \nu_{\theta,4,2}\vecd_6^\ast]},\\
&\CT_{\idp,\tp}^{(\SFp)}:=\left\{\sCp:=\msgp\cdot (g_T^{\alpha})^z, \sCp_0:=g_1^{z( \vecd_1 + \idp \vecd_2 +\tp\vecd_3 ) + [\chi_4( \vecd_4 + \idp \vecd_5 +\tp \vecd_6 )] }\right\},
\end{align*}
where $\alpha_{\theta},r_{\theta,1},\nu_{\theta,4,1},r_{\theta,2},\nu_{\theta,4,2},z,\chi_4\getsr \Z_q$.

Note that nominally semi-functional pair of Type I is used to transform the non-challenge private key and key update queries into semi-functional ones while Type II is for the challenge private key and key update queries.

Assume that a probabilistic polynomial-time adversary $\AdvA$ makes at most $\qnop$ private key queries $\idp_1,\ldots, \idp_{\qnop}$ and $\qntp$ key update queries $\tp_1,\ldots, \tp_{\qntp}$.
Since there are many types of adversaries according to whether the challenges $\idp_{(0)}^\ast$, $\idp_{(1)}^\ast$ $\tp_{(0)}^\ast$, $\tp_{(1)}^\ast$ being queried and the restriction of queries, in order to simplify and unify reduction, we add four dumb queries $\idp_{\qnop+1},\idp_{\qnop+2},\tp_{\qntp+1},\tp_{\qntp+2}$ (the keys for these queries will not be given to $\AdvA$),
which makes the challenge identities $\idp_{(0)}^\ast,\idp_{(1)}^\ast$ and times $\tp_{(0)}^\ast,\tp_{(1)}^\ast$ be included in the $\qnop+2$ private key queries and the the $\qntp+2$ key update queries.
For any adversary, we use values $\varphi_1,\varphi_2$ ($0<\varphi_1<\varphi_2<\qnop+2$) to indicate the positions of $\idp_{(0)}^\ast,\idp_{(1)}^\ast$ being queried, namely either the $\varphi_1$-th
or $\varphi_2$-th query is $\idp_{(0)}^\ast$ and the other is $\idp_{(1)}^\ast$. Similarly, we use values $\varphi_3,\varphi_4$ ($0<\varphi_3<\varphi_4<\qntp+2$) to indicate the positions of $\tp_{(0)}^\ast,\tp_{(1)}^\ast$ being queried.

Our proof of security consists of the following sequence of games between the adversary $\AdvA$ and challengers.
\begin{IEEEitemize}
\item
$\Gm_{Real}$: is the real security game.

\item
$\Gm_{Real'}$:
is a preliminary game, which is the same as $\Gm_{Real}$ except that the challenger picks $\phi_1,\phi_2\getsr[\qnop+2]$ ($0<\phi_1<\phi_2<\qnop+2$) and $\phi_3,\phi_4\getsr[\qntp+2]$ ($0<\phi_3<\phi_4<\qntp+2$) before
setup, and the game is aborted if $\phi_i\neq \varphi_i$ for any $i\in[4]$.

//Guess the positions of the challenge identities $\idp_{(0)}^\ast,\idp_{(1)}^\ast$ and times $\tp_{(0)}^\ast,\tp_{(1)}^\ast$. If the guess is incorrect then the game aborts.
Re-write
\begin{align*}
&\Gamma_1:=\{\idp_1',\ldots,\idp_{\qnop}'\}=\{\idp_1,\ldots,\idp_{\qnop+2}\}\backslash\{\idp_{\varphi_1},\idp_{\varphi_2}\}\\
&\Gamma_2:=\{\tp_1',\ldots,\tp_{\qntp}'\}=\{\tp_1,\ldots,\tp_{\qntp+2}\}\backslash\{\tp_{\varphi_3},\tp_{\varphi_4}\}.
\end{align*}

\item $\Gm_{0}$: is the same as $\Gm_{Real'}$ except that the challenge ciphertext is semi-functional.

\item $\Gm_{\kappa_1,\kappa_2}$: for $\kappa_1$ from $1$ to $\qnop$, for $\kappa_2$ from $0$ to $\lceil\log{N_{max}}\rceil$, $\Gm_{\kappa_1,\kappa_2}$ is the same as $\Gm_{0}$ except that the first $\kappa_1-1$ private keys and the first $\kappa_2$ components of the $\kappa_1$-th private key for $\Gamma_1$ are
semi-functional and the remaining keys are normal.

//Transform all private keys into semi-functional ones (one by one and node by node) except the $\phi_1$-th
and $\phi_2$-th private queries. Namely, the private keys for the challenge identities $\idp_{(0)}^\ast,\idp_{(1)}^\ast$ (if queried) are still normal.
Note that the number of nodes associated with a private key is $\lceil\log{N_{max}}\rceil$. Moreover $\Gm_{1,0}$ and $\Gm_{0}$, $\Gm_{\kappa_1,\lceil\log{N_{max}}\rceil}$ and $\Gm_{\kappa_1+1,0}$ are identical.

\item $\Gm_{\kappa_1,\kappa_2}$: for $\kappa_1$ from $\qnop+1$ to $\qntp$, for $\kappa_2$ from $0$ to $N_{max}$, $\Gm_{\kappa_1,\kappa_2}$ is the same as $\Gm_{\qnop, \lceil\log{N_{max}}\rceil}$ (namely all private keys for $\Gamma_1$ are semi-functional) except that the first $\kappa_1-\qnop-1$ key updates and the first $\kappa_2$ components of the $(\kappa_1-\qnop)$-th key update for $\Gamma_2$ are
semi-functional and the remaining key updates are normal.

//Transform all key updates into semi-functional ones (one by one and node by node) except the $\phi_3$-th
and $\phi_4$-th key update queries. Namely, the key updates for the challenge times $\tp_{(0)}^\ast,\tp_{(1)}^\ast$ (if queried) are still normal.
Note that a key update for a time updates at most $N_{max}$ nodes.
Moreover, $\Gm_{\qnop,\lceil\log{N_{max}}\rceil}$ and $\Gm_{\qnop+1,0}$, $\Gm_{\kappa_1,N_{max}}$ and $\Gm_{\kappa_1+1,0}$ are identical.

\item $\Gm_{\qnop+\qntp+1,\kappa_2}$: for $\kappa_2$ from $0$ to $4N_{max}$, $\Gm_{\qnop+\qntp+1,\kappa_2}$ is the same as $\Gm_{\qnop+\qntp, N_{max}}$ (namely all private keys for $\Gamma_1$ and key updates for $\Gamma_2$ are semi-functional) except that the $\phi_1,\phi_2$-th private keys, the $\phi_3,\phi_4$-th key updates for the first $\kappa_2$ nodes are
semi-functional and the remaining keys are normal.

//Transform the $\varphi_1,\varphi_2$-th private key and the $\varphi_3,\varphi_4$-th key update queries into semi-functional ones (node by node).
Note that there are at most $2^{\lceil\log{N_{max}}\rceil}$ ($\leq 4N_{max}$) nodes in the binary tree.
Moreover, $\Gm_{\qnop+\qntp,N_{max}}$ and $\Gm_{\qnop+\qntp+1,0}$ are identical, namely all keys are
semi-functional in $\Gm_{\qnop+\qntp+1,4N_{max}}$.

\item $\Gm_{Final}$: is the same as $\Gm_{\qnop+\qntp+1,4N_{max}}$,
except that the challenge ciphertext is
a semi-functional encryption of a random message in $G_T$ and under a random identity in $\Z_q$ a random time in  $\Z_q$.
We denote the challenge ciphertext in $\Gm_{Final}$ as $\CT_{\idp_{(\Rp)},\tp_{(\Rp)}}^{(\Rp)}$.

\end{IEEEitemize}

We prove the following lemmas to show the above games are
indistinguishable.
The advantage gap between $\Gm_{Real}$ and
$\Gm_{0}$ is bounded by the advantage of the Subspace assumption in
$G_1$. Additionally, we require a statistical indistinguishability
argument to show that the distribution of the challenge ciphertext
remains the same from the adversary's view. Similarly, the advantage
gap between any two consecutive games of $\Gm_{1,1}$ to
$\Gm_{\qnop+\qntp+1,4N_{max}}$ is bounded by the advantage of Subspace
assumption in $G_2$. Finally, we statistically transform
$\Gm_{\qnop+\qntp+1,4N_{max}}$ to $\Gm_{Final}$ in one step, i.e., we show
the joint distributions of parameters in these two games are
equivalent from the adversary's view.

We let $\Adv_{\AdvA}^{\Gm_{Real}}$ denote an adversary $\AdvA$'s advantage in the real game.

\begin{mylem} \label{lem:security-ribe-1}
For any adversary $\AdvA$, $\Adv_{\AdvA}^{\Gm_{Real}}(\lambda)\leq (\qnop \qntp)^2 \cdot\Adv_{\AdvA}^{\Gm_{Real'}}(\lambda)$.
\end{mylem}
\begin{IEEEproof}
Since $\phi_1,\phi_2,\phi_3,\phi_4$ are uniformly and independently
generated, which are hidden from the adversary $\AdvA$'s view. The
game is non-aborted with probability
\[
\frac{4}{(\qnop+2)(\qnop+1)(\qntp+2)(\qntp+1)}.
\]
Thus,
\begin{align*}
&\Adv_{\AdvA}^{\Gm_{Real}}(\lambda) = \frac{(\qnop+2)(\qnop+1)(\qntp+2)(\qntp+1)}{4}\cdot\Adv_{\AdvA}^{\Gm_{Real'}}(\lambda)\\
&\quad\quad\quad\quad\quad\quad \leq (\qnop \qntp)^2 \cdot\Adv_{\AdvA}^{\Gm_{Real'}}(\lambda).
\end{align*}
\end{IEEEproof}

\begin{mylem}\label{lem:security-ribe-2}
Suppose that there exists an adversary $\AdvA$ where $|\Adv_{\AdvA}^{\Gm_{Real'}}(\lambda)-\Adv_{\AdvA}^{\Gm_{0}}(\lambda)|=\epsilon$. Then there exists an algorithm $\AdvB_0$ such that $\Adv_{\AdvB_0}^{\DSop}(\lambda)=\epsilon + \frac{2}{q}$, with $K=3$ and $N=6$.
\end{mylem}
\begin{IEEEproof}
$\AdvB_0$ is given
\[
D:=(\mathbb{G}; g_2^{\vecb_1^\ast},g_2^{\vecb_2^\ast},g_2^{\vecb_3^\ast},g_1^{\vecb_1},\ldots,g_1^{\vecb_6},U_1,U_2,U_3,\mu_2).
\]
along with $T_1, T_2, T_3$. We require that $\AdvB_0$ decides
whether $T_1, T_2, T_3$ are distributed as
$g_1^{\tau_1\vecb_1},g_1^{\tau_1\vecb_2},g_1^{\tau_1\vecb_3}$ or
$g_1^{\tau_1\vecb_1+\tau_2\vecb_4},g_1^{\tau_1\vecb_2+\tau_2\vecb_5},g_1^{\tau_1\vecb_3+\tau_2\vecb_6}$.

$\AdvB_0$ simulates $\Gm_{Real'}$ or $\Gm_0$ with $\AdvA$, depending
on the distribution of $T_1, T_2, T_3$. To compute the public
parameters and master key, $\AdvB_0$ chooses a random
invertible matrix $\matA \in \Z_q^{3\times 3}$ ($\matA$ is
invertible with overwhelming probability if it is uniformly picked) and implicitly sets dual orthonormal bases $\Ddual,\Ddual^\ast$ to:
\begin{align*}
&\vecd_1:=\vecb_1, \quad\vecd_2:=\vecb_2, \quad\vecd_3:=\vecb_3, \quad (\vecd_4,\vecd_5,\vecd_6):=(\vecb_4,\vecb_5,\vecb_6)\matA,\\
&\vecd_1^\ast:=\vecb_1^\ast, \quad\vecd_2^\ast:=\vecb_2^\ast, \quad\vecd_3^\ast:=\vecb_3^\ast, \quad (\vecd_4^\ast,\vecd_5^\ast,\vecd_6^\ast):=(\vecb_4^\ast,\vecb_5^\ast,\vecb_6^\ast)(\matA^{-1})^\tran.
\end{align*}
\noindent We note that $\Ddual,\Ddual^\ast$ are properly
distributed, and reveal no information about $\matA$. Moreover,
$\AdvB_0$ cannot generate
$g_2^{\vecd_4^\ast},g_2^{\vecd_5^\ast},g_2^{\vecd_6^\ast}$, but
these will not be needed for creating normal private keys and key
updates. $\AdvB_0$ chooses random value $\alpha\in \Z_q$ and
computes $g_T^\alpha:=e(g_1, g_2)^{\alpha\vecd_1\cdot\vecd_1^\ast}$. It then
gives $\AdvA$ the public parameters
\[
\PP:=\{\mathbb{G}; g_T^\alpha,g_1^{\vecd_1},g_1^{\vecd_2},g_1^{\vecd_3}\}.
\]
\noindent The master key
\[
\MK:=\{\alpha,g_2^{\vecd_1^\ast},g_2^{\vecd_2^\ast},g_2^{\vecd_3^\ast}\}
\]
is known to $\AdvB_0$, which allows $\AdvB_0$ to respond to all of
$\AdvA$'s queries by calling the normal private keys, key updates,
and key revocation algorithms.

$\AdvA$ sends $\AdvB_0$ two pairs
$(\idp_{(0)}^\ast,\tp_{(0)}^\ast,\msgp_{(0)}^\ast)$ and
$(\idp_{(1)}^\ast,\tp_{(1)}^\ast,\msgp_{(1)}^\ast)$. $\AdvB_0$
chooses a random bit $\beta \in\{0, 1\}$ and encrypts
$\msgp^\ast_{(\beta)}$ under $(\idp^\ast_{(\beta)},\tp^\ast_{(\beta)})$
as follows:
\[
\sCp:=\msgp^\ast_{(\beta)}\cdot\left( e(T_1,g_2^{\vecb_1^\ast})\right)^{\alpha}=\msgp^\ast_{(\beta)}\cdot(g_T^\alpha)^z,
\quad \sCp_0:=T_1(T_2)^{\idp^\ast_{(\beta)}}(T_3)^{\tp^\ast_{(\beta)}},
\]
\noindent where $\AdvB_0$ has implicitly set $z:=\tau_1$. It gives
the ciphertext $(\sCp,\sCp_0)$ to $\AdvA$.

Now, if $T_1, T_2, T_3$ are equal to
$g_1^{\tau_1\vecb_1},g_1^{\tau_1\vecb_2},g_1^{\tau_1\vecb_3}$, then
this is a properly distributed normal encryption of
$\msgp^\ast_{(\beta)}$. In this case, $\AdvB_0$ has properly simulated
$\Gm_{Real'}$. If $T_1, T_2,T_3$ are equal to
$g_1^{\tau_1\vecb_1+\tau_2\vecb_4},g_1^{\tau_1\vecb_2+\tau_2\vecb_5},g_1^{\tau_1\vecb_3+\tau_2\vecb_6}$
instead, then the ciphertext element $\sCp_0$ has an additional term
of
\[
\tau_2\vecb_4+\idp^\ast_{(\beta)}\tau_2\vecb_5+\tp^\ast_{(\beta)}\tau_2\vecb_6
\]
\noindent in its exponent. The coefficients here in the basis
$\vecb_4,\vecb_5,\vecb_6$ form the vector
$(\tau_2,\idp^\ast_{(\beta)}\tau_2,\tp^\ast_{(\beta)}\tau_2)$. To
compute the coefficients in the basis $\vecd_4,\vecd_5,\vecd_6$, we
multiply the matrix $\matA^{-1}$ by the transpose of this vector,
obtaining
$\tau_2\matA^{-1}(1,\idp^\ast_{(\beta)},\tp^\ast_{(\beta)})^\tran$.
Since $\matA$ is random (everything else given to $\AdvA$ has been distributed
independently of $\matA$), these coefficients are uniformly random except with probability $2/q$ (namely, the cases $\tau_2$ defined in Subspace problem is zero, $(\chi_4,\chi_5,\chi_6)$ defined in Equation~\ref{equ:semi-ct-ribe} is the zero vector
) from Lemma~\ref{lem:mat-indist}.
Therefore, in
this case, $\AdvB_0$ has properly simulated $\Gm_0$. This allows
$\AdvB_0$ to leverage $\AdvA$'s advantage $\epsilon$ between
$\Gm_{Real'}$ and $\Gm_0$ to achieve an advantage $\epsilon +
\frac{2}{q}$ against the Subspace assumption in $G_1$, namely
$\Adv_{\AdvB_0}^{\DSop}(\lambda)=\epsilon +
\frac{2}{q}$.
\end{IEEEproof}

\begin{mylem}\label{lem:security-ribe-3}
For $\kappa_1$ from $1$ to $\qnop$, for $\kappa_2$ from $0$ to $\lceil\log{N_{max}}\rceil$, suppose that there exists an adversary $\AdvA$ where $|\Adv_{\AdvA}^{\Gm_{\kappa_1,\kappa_2-1}}(\lambda)-\Adv_{\AdvA}^{\Gm_{\kappa_1,\kappa_2}}(\lambda)|=\epsilon$.
Then there exists an algorithm $\AdvB_{\kappa_1,\kappa_2}$ such that
$\Adv_{\AdvB_{\kappa_1,\kappa_2}}^{\DStp}(\lambda)=\epsilon
+ \frac{6}{q} $, with $K=3$ and $N=6$.
\end{mylem}
\begin{IEEEproof}
$\AdvB_{\kappa_1,\kappa_2}$ is given
\[
D:=(\mathbb{G};g_1^{\vecb_1},g_1^{\vecb_2},g_1^{\vecb_3},g_2^{\vecb_1^\ast},\ldots,g_2^{\vecb_6^\ast},U_1,U_2,U_3,\mu_2)
\]
along with $T_1, T_2,T_3$. We require that
$\AdvB_{\kappa_1,\kappa_2}$ decides whether $T_1, T_2, T_3$ are
distributed as
$g_2^{\tau_1\vecb_1^\ast},g_2^{\tau_1\vecb_2^\ast},g_2^{\tau_1\vecb_3^\ast}$
or
$g_2^{\tau_1\vecb_1^\ast+\tau_2\vecb_4^\ast},g_2^{\tau_1\vecb_2^\ast+\tau_2\vecb_5^\ast},g_2^{\tau_1\vecb_3^\ast+\tau_2\vecb_6^\ast}$.

$\AdvB_{\kappa_1,\kappa_2}$ simulates $\Gm_{\kappa_1,\kappa_2}$ or
$\Gm_{\kappa_1,\kappa_2-1}$ with $\AdvA$, depending on the
distribution of $T_1, T_2, T_3$. To compute the public parameters
and master key, $\AdvB_{\kappa_1,\kappa_2}$ chooses a random matrix
$\matA \in \Z_q^{3\times 3}$ (with all but negligible probability,
$\matA$ is invertible). We then implicitly set dual orthonormal
bases $\Ddual,\Ddual^\ast$ to:
\begin{align*}
&\vecd_1:=\vecb_1, \quad\vecd_2:=\vecb_2, \quad\vecd_3:=\vecb_3,\quad (\vecd_4,\vecd_5,\vecd_6):=(\vecb_4,\vecb_5,\vecb_6)\matA,\\
&\vecd_1^\ast:=\vecb_1^\ast, \quad\vecd_2^\ast:=\vecb_2^\ast, \quad\vecd_3^\ast:=\vecb_3^\ast,
\quad (\vecd_4^\ast,\vecd_5^\ast,\vecd_6^\ast):=(\vecb_4^\ast,\vecb_5^\ast,\vecb_6^\ast)(\matA^{-1})^\tran.
\end{align*}
\noindent We note that $\Ddual,\Ddual^\ast$ are properly
distributed, and reveal no information about $\matA$.
$\AdvB_{\kappa_1,\kappa_2}$ chooses random value $\alpha\in\Z_q$ and
compute $g_T^\alpha:=e(g_1, g_2)^{\alpha\vecd_1\cdot\vecd_1^\ast}$. $\AdvB$ can
gives $\AdvA$ the public parameters
\[
\PP:=\{\mathbb{G}; g_T^\alpha,g_1^{\vecd_1},g_1^{\vecd_2},g_1^{\vecd_3}\}.
\]
\noindent The master key
\[
\MK:=\{\alpha,g_2^{\vecd_1^\ast},g_2^{\vecd_2^\ast},g_2^{\vecd_3^\ast}\}
\]
is known to $\AdvB_{\kappa_1,\kappa_2}$, which allows
$\AdvB_{\kappa_1,\kappa_2}$ to respond to all of $\AdvA$'s private
key and key update queries by calling the normal key generation
algorithm. Since $\AdvB_{\kappa_1,\kappa_2}$ also knows
$g_2^{\vecd_4^\ast}$, $g_2^{\vecd_5^\ast}$, and
$g_2^{\vecd_6^\ast}$, it can easily produce semi-functional keys. To
answer the key queries that $\AdvA$ makes,
$\AdvB_{\kappa_1,\kappa_2}$ runs the semi-functional private key and
key update generation algorithm to produce semi-functional keys and
gives these to $\AdvA$. To answer the $\kappa_2$-th component of the
$\kappa_1$-th private key for $\idp_{\kappa_1}'$,
$\AdvB_{\kappa_1,\kappa_2}$ responds with:
\[
\sKp_{\idp_{\kappa_1}',\theta}:=(g_2^{\vecb_1^\ast})^{\alpha_{\theta,1}} T_1^{\idp_{\kappa_1}'}(T_2)^{-1}.
\]
\noindent This implicitly sets $r_{\theta,1}:=\tau_1$. If $T_1, T_2,
T_3$ are equal to
$g_2^{\tau_1\vecb_1^\ast},g_2^{\tau_1\vecb_2^\ast},g_2^{\tau_1\vecb_3^\ast}$,
then this is a properly distributed normal private key. Otherwise, if $T_1,
T_2, T_3$ are equal to
$g_2^{\tau_1\vecb_1^\ast+\tau_2\vecb_4^\ast},g_2^{\tau_1\vecb_2^\ast+\tau_2\vecb_5^\ast},g_2^{\tau_1\vecb_3^\ast+\tau_2\vecb_6^\ast}$,
then this is a semi-functional key, whose exponent vector includes
\begin{align}\label{equ1-ribe}
\idp_{\kappa_1}'\tau_2\vecb_4^\ast-\tau_2\vecb_5^\ast
\end{align}
as its component in the span of
$\vecb_4^\ast,\vecb_5^\ast,\vecb_6^\ast$. To respond to the
remaining key queries, $\AdvB_{\kappa_1,\kappa_2}$ simply runs the
normal key generation algorithm.

At some point, $\AdvA$ sends $\AdvB_{\kappa_1,\kappa_2}$
two pairs $(\idp_{(0)}^\ast,\tp_{(0)}^\ast,\msgp_{(0)}^\ast)$ and
$(\idp_{(1)}^\ast,\tp_{(1)}^\ast,\msgp_{(1)}^\ast)$. $\AdvB_0$
chooses a random bit $\beta \in\{0, 1\}$ and encrypts
$\msgp^\ast_{(\beta)}$ under $(\idp^\ast_{(\beta)},\tp^\ast_{(\beta)})$
as follows:
\[
\sCp:=\msgp^\ast_{(\beta)}\cdot\left( e(U_1,g_2^{\vecb_1^\ast})\right)^{\alpha}=\msgp^\ast_{(\beta)}\cdot (g_T^\alpha)^z,
\quad \sCp_0:=U_1(U_2)^{\idp^\ast_{(\beta)}}(U_3)^{\tp^\ast_{(\beta)}},
\]
\noindent where $\AdvB_{\kappa_1,\kappa_2}$ has implicitly set
$z:=\mu_1$. The ``semi-functional part'' of the exponent vector here
is:
\begin{align}\label{equ2-ribe}
\mu_2\vecb_4+\idp^\ast_{(\beta)}\mu_2\vecb_5+\tp^\ast_{(\beta)}\mu_2\vecb_6.
\end{align}
We observe that if $\idp^\ast_{(\beta)} = \idp_{\kappa_1}'$ (which
is impossible), then vectors~\ref{equ1-ribe} and~\ref{equ2-ribe} would be
orthogonal, resulting in a nominally semi-functional ciphertext and
key pair ($\AdvB_{\kappa_1,\kappa_2}$ can also use $T_1, T_2,T_3$ to
generate private key part for $\tp^\ast_{(\beta)}$) of Type I. It
gives the ciphertext
$(\sCp,\sCp_0)$ to $\AdvA$.

We now argue that since $\idp^\ast_{(\beta)}\neq \idp_{\kappa_1}'$,
in $\AdvA$'s view the vectors~\ref{equ1-ribe} and~\ref{equ2-ribe} are
distributed as random vectors in the spans of
$\vecd_4^\ast,\vecd_5^\ast,\vecd_6^\ast$ and
$\vecd_4,\vecd_5,\vecd_6$ respectively. To see this, we take the
coefficients of vectors~\ref{equ1-ribe} and~\ref{equ2-ribe} in terms of the
bases $\vecb_4^\ast,\vecb_5^\ast,\vecb_6^\ast$ and
$\vecb_4,\vecb_5,\vecb_6$ respectively and translate them into
coefficients in terms of the bases
$\vecd_4^\ast,\vecd_5^\ast,\vecd_6^\ast$ and
$\vecd_4,\vecd_5,\vecd_6$. Using the change of basis matrix $\matA$,
we obtain the new coefficients (in vector form) as:
\[
\tau_2\matA^\tran(\idp_{\kappa_1}',-1,0)^\tran,\, \mu_2\matA^{-1}(1,\idp^\ast_{(\beta)},\tp^\ast_{(\beta)})^\tran.
\]
Since the distribution of everything given to $\matA$ except for the
$\kappa_2$-th component of the $\kappa_1$-th private key
$\sKp_{\idp_{\kappa_1}',\theta}$ and the challenge ciphertext
$(\sCp,\sCp_0)$ is independent
of the random matrix $\matA$ and $\idp^\ast_{(\beta)}\neq
\idp_{\kappa_1}'$,
we can conclude
that these coefficients are uniformly except with probability $4/q$ (namely, the cases $\mu_2$ or $\tau_2$ defined in Subspace problem is zero, $(\chi_4,\chi_5,\chi_6)$ or $(\nu_{\theta,4,1},\nu_{\theta,5,1},\nu_{\theta,6,1})$ defined in Equations~\ref{equ:semi-ct-ribe} and~\ref{equ:semi-key-ribe-1} is the zero vector) from Lemma~\ref{lem:mat-indist}. Thus, $\AdvB_{\kappa_1,\kappa_2}$ has
properly simulated $\Gm_{\kappa_1,\kappa_2}$ in this case.

If $T_1,T_2, T_3$ are equal to
$g_2^{\tau_1\vecb_1^*},g_2^{\tau_1\vecb_2^*},g_2^{\tau_1\vecb_3^*}$,
then the coefficients of the vector~\ref{equ2-ribe} are uniformly except with probability $2/q$ (namely, the cases $\mu_2$ defined in Subspace problem is zero, $(\chi_4,\chi_5,\chi_6)$ defined in Equations~\ref{equ:semi-ct-ribe} is the zero vector) from Lemma~\ref{lem:mat-indist}. Thus, $\AdvB_{\kappa_1,\kappa_2}$ has
properly simulated $\Gm_{\kappa_1,\kappa_2-1}$ in this case.

In summary, $\AdvB_{\kappa_1,\kappa_2}$ has properly simulated
either $\Gm_{\kappa_1,\kappa_2-1}$ or $\Gm_{\kappa_1,\kappa_2}$ for
$\AdvA$, depending on the distribution of $T_1, T_2,T_3$. It can
therefore leverage $\AdvA$'s advantage $\epsilon$ between these
games to obtain an advantage $\epsilon + \frac{6}{q}$ against
the Subspace assumption in $G_2$, namely
$\Adv_{\AdvB_\kappa}^{\DStp}(\lambda)=\epsilon +
\frac{6}{q}$.
\end{IEEEproof}

\begin{mylem}\label{lem:security-ribe-4}
For $\kappa_1$ from $\qnop+1$ to $\qnop+\qntp$, for $\kappa_2$ from $0$ to $N_{max}$, suppose that there exists an adversary $\AdvA$ where $|\Adv_{\AdvA}^{\Gm_{\kappa_1,\kappa_2-1}}(\lambda)-\Adv_{\AdvA}^{\Gm_{\kappa_1,\kappa_2}}(\lambda)|=\epsilon$.
Then there exists an algorithm $\AdvB_{\kappa_1,\kappa_2}$ such that $\Adv_{\AdvB_{\kappa_1,\kappa_2}}^{\DStp}(\lambda)=\epsilon + \frac{6}{q}$, with $K=3$ and $N=6$.
\end{mylem}
\begin{IEEEproof}
$\AdvB_{\kappa_1,\kappa_2}$ is given
\[
D:=(\mathbb{G};g_1^{\vecb_1},g_1^{\vecb_2},g_1^{\vecb_3},g_2^{\vecb_1^\ast},\ldots,g_2^{\vecb_6^\ast},U_1,U_2,U_3,\mu_2)
\]
along with $T_1, T_2,T_3$. We require that
$\AdvB_{\kappa_1,\kappa_2}$ decides whether $T_1, T_2, T_3$ are
distributed as
$g_2^{\tau_1\vecb_1^\ast},g_2^{\tau_1\vecb_2^\ast},g_2^{\tau_1\vecb_3^\ast}$
or
$g_2^{\tau_1\vecb_1^\ast+\tau_2\vecb_4^\ast},g_2^{\tau_1\vecb_2^\ast+\tau_2\vecb_5^\ast},g_2^{\tau_1\vecb_3^\ast+\tau_2\vecb_6^\ast}$.

$\AdvB_{\kappa_1,\kappa_2}$ simulates $\Gm_{\kappa_1,\kappa_2}$ or
$\Gm_{\kappa_1,\kappa_2-1}$ with $\AdvA$, depending on the
distribution of $T_1, T_2, T_3$. To compute the public parameters
and master key, $\AdvB_{\kappa_1,\kappa_2}$ chooses a random matrix
$\matA \in \Z_q^{3\times 3}$ (with all but negligible probability,
$\matA$ is invertible). We then implicitly set dual orthonormal
bases $\Ddual,\Ddual^\ast$ to:
\begin{align*}
&\vecd_1:=\vecb_1, \quad\vecd_2:=\vecb_2, \quad\vecd_3:=\vecb_3,\quad (\vecd_4,\vecd_5,\vecd_6):=(\vecb_4,\vecb_5,\vecb_6)\matA,\\
&\vecd_1^\ast:=\vecb_1^\ast, \quad\vecd_2^\ast:=\vecb_2^\ast, \quad\vecd_3^\ast:=\vecb_3^\ast,
\quad (\vecd_4^\ast,\vecd_5^\ast,\vecd_6^\ast):=(\vecb_4^\ast,\vecb_5^\ast,\vecb_6^\ast)(\matA^{-1})^\tran.
\end{align*}
\noindent We note that $\Ddual,\Ddual^\ast$ are properly
distributed, and reveal no information about $\matA$.
$\AdvB_{\kappa_1,\kappa_2}$ chooses random value $\alpha\in\Z_q$ and
compute $g_T^\alpha:=e(g_1, g_2)^{\alpha\vecd_1\cdot\vecd_1^\ast}$. $\AdvB$ can
gives $\AdvA$ the public parameters
\[
\PP:=\{\mathbb{G}; g_T^\alpha,g_1^{\vecd_1},g_1^{\vecd_2},g_1^{\vecd_3}\}.
\]
\noindent The master key
\[
\MK:=\{\alpha,g_2^{\vecd_1^\ast},g_2^{\vecd_2^\ast},g_2^{\vecd_3^\ast}\}
\]
is known to $\AdvB_{\kappa_1,\kappa_2}$, which allows
$\AdvB_{\kappa_1,\kappa_2}$ to respond to all of $\AdvA$'s private
key and key update queries by calling the normal key generation
algorithm. Since $\AdvB_{\kappa_1,\kappa_2}$ also knows
$g_2^{\vecd_4^\ast}$, $g_2^{\vecd_5^\ast}$, and
$g_2^{\vecd_6^\ast}$, it can easily produce semi-functional keys. To
answer the key queries that $\AdvA$ makes,
$\AdvB_{\kappa_1,\kappa_2}$ runs the semi-functional private key and
key update generation algorithm to produce semi-functional keys and
gives these to $\AdvA$. To answer the $\kappa_2$-th component of the
$\kappa_1$-th private key for $\idp_{\kappa_1}'$,
$\AdvB_{\kappa_1,\kappa_2}$ responds with:
\[
\sKp_{\idp_{\kappa_1}',\theta}:=(g_2^{\vecb_1^\ast})^{\alpha_{\theta,1}} T_1^{\idp_{\kappa_1}'}(T_2)^{-1}.
\]
\noindent This implicitly sets $r_{\theta,1}:=\tau_1$. If $T_1, T_2,
T_3$ are equal to
$g_2^{\tau_1\vecb_1^\ast},g_2^{\tau_1\vecb_2^\ast},g_2^{\tau_1\vecb_3^\ast}$,
then this is a properly distributed normal private key. Otherwise, if $T_1,
T_2, T_3$ are equal to
$g_2^{\tau_1\vecb_1^\ast+\tau_2\vecb_4^\ast},g_2^{\tau_1\vecb_2^\ast+\tau_2\vecb_5^\ast},g_2^{\tau_1\vecb_3^\ast+\tau_2\vecb_6^\ast}$,
then this is a semi-functional key, whose exponent vector includes
\begin{align}\label{equ11-ribe}
\idp_{\kappa_1}'\tau_2\vecb_4^\ast-\tau_2\vecb_5^\ast
\end{align}
as its component in the span of
$\vecb_4^\ast,\vecb_5^\ast,\vecb_6^\ast$. To respond to the
remaining key queries, $\AdvB_{\kappa_1,\kappa_2}$ simply runs the
normal key generation algorithm.

At some point, $\AdvA$ sends $\AdvB_{\kappa_1,\kappa_2}$
two pairs $(\idp_{(0)}^\ast,\tp_{(0)}^\ast,\msgp_{(0)}^\ast)$ and
$(\idp_{(1)}^\ast,\tp_{(1)}^\ast,\msgp_{(1)}^\ast)$. $\AdvB_0$
chooses a random bit $\beta \in\{0, 1\}$ and encrypts
$\msgp^\ast_{(\beta)}$ under $(\idp^\ast_{(\beta)},\tp^\ast_{(\beta)})$
as follows:
\[
\sCp:=\msgp^\ast_{(\beta)}\cdot\left( e(U_1,g_2^{\vecb_1^\ast})\right)^{\alpha}=\msgp^\ast_{(\beta)}\cdot (g_T^\alpha)^z,
\quad \sCp_0:=U_1(U_2)^{\idp^\ast_{(\beta)}}(U_3)^{\tp^\ast_{(\beta)}},
\]
\noindent where $\AdvB_{\kappa_1,\kappa_2}$ has implicitly set
$z:=\mu_1$. The ``semi-functional part'' of the exponent vector here
is:
\begin{align}\label{equ22-ribe}
\mu_2\vecb_4+\idp^\ast_{(\beta)}\mu_2\vecb_5+\tp^\ast_{(\beta)}\mu_2\vecb_6.
\end{align}
We observe that if $\idp^\ast_{(\beta)} = \idp_{\kappa_1}'$ (which
is impossible), then vectors~\ref{equ11-ribe} and~\ref{equ22-ribe} would be
orthogonal, resulting in a nominally semi-functional ciphertext and
key pair ($\AdvB_{\kappa_1,\kappa_2}$ can also use $T_1, T_2,T_3$ to
generate private key part for $\tp^\ast_{(\beta)}$) of Type I. It
gives the ciphertext
$(\sCp,\sCp_0)$ to $\AdvA$.

We now argue that since $\idp^\ast_{(\beta)}\neq \idp_{\kappa_1}'$,
in $\AdvA$'s view the vectors~\ref{equ11-ribe} and~\ref{equ22-ribe} are
distributed as random vectors in the spans of
$\vecd_4^\ast,\vecd_5^\ast,\vecd_6^\ast$ and
$\vecd_4,\vecd_5,\vecd_6$ respectively. To see this, we take the
coefficients of vectors~\ref{equ11-ribe} and~\ref{equ22-ribe} in terms of the
bases $\vecb_4^\ast,\vecb_5^\ast,\vecb_6^\ast$ and
$\vecb_4,\vecb_5,\vecb_6$ respectively and translate them into
coefficients in terms of the bases
$\vecd_4^\ast,\vecd_5^\ast,\vecd_6^\ast$ and
$\vecd_4,\vecd_5,\vecd_6$. Using the change of basis matrix $\matA$,
we obtain the new coefficients (in vector form) as:
\[
\tau_2\matA^\tran(\idp_{\kappa_1}',-1,0)^\tran,\, \mu_2\matA^{-1}(1,\idp^\ast_{(\beta)},\tp^\ast_{(\beta)})^\tran.
\]
Since the distribution of everything given to $\matA$ except for the
$\kappa_2$-th component of the $\kappa_1$-th private key
$\sKp_{\idp_{\kappa_1}',\theta}$ and the challenge ciphertext
$(\sCp,\sCp_0)$ is independent
of the random matrix $\matA$ and $\idp^\ast_{(\beta)}\neq
\idp_{\kappa_1}'$,
we can conclude
that these coefficients are uniformly except with probability $4/q$ (namely, the cases $\mu_2$ or $\tau_2$ defined in Subspace problem is zero, $(\chi_4,\chi_5,\chi_6)$ or $(\nu_{\theta,4,1},\nu_{\theta,5,1},\nu_{\theta,6,1})$ defined in Equations~\ref{equ:semi-ct-ribe} and~\ref{equ:semi-key-ribe-1} is the zero vector) from Lemma~\ref{lem:mat-indist}. Thus, $\AdvB_{\kappa_1,\kappa_2}$ has
properly simulated $\Gm_{\kappa_1,\kappa_2}$ in this case.

If $T_1,T_2, T_3$ are equal to
$g_2^{\tau_1\vecb_1^*},g_2^{\tau_1\vecb_2^*},g_2^{\tau_1\vecb_3^*}$,
then the coefficients of the vector~\ref{equ22-ribe} are uniformly except with probability $2/q$ (namely, the cases $\mu_2$ defined in Subspace problem is zero, $(\chi_4,\chi_5,\chi_6)$ defined in Equations~\ref{equ:semi-ct-ribe} is the zero vector) from Lemma~\ref{lem:mat-indist}. Thus, $\AdvB_{\kappa_1,\kappa_2}$ has
properly simulated $\Gm_{\kappa_1,\kappa_2-1}$ in this case.

In summary, $\AdvB_{\kappa_1,\kappa_2}$ has properly simulated
either $\Gm_{\kappa_1,\kappa_2-1}$ or $\Gm_{\kappa_1,\kappa_2}$ for
$\AdvA$, depending on the distribution of $T_1, T_2,T_3$. It can
therefore leverage $\AdvA$'s advantage $\epsilon$ between these
games to obtain an advantage $\epsilon + \frac{6}{q}$ against
the Subspace assumption in $G_2$, namely
$\Adv_{\AdvB_\kappa}^{\DStp}(\lambda)=\epsilon +
\frac{6}{q}$.
\\
\\
$\AdvB_{\kappa_1,\kappa_2}$ is given
\[
D:=(\mathbb{G};g_1^{\vecb_1},g_1^{\vecb_2},g_1^{\vecb_3},g_2^{\vecb_1^\ast},\ldots,g_2^{\vecb_6^\ast},U_1,U_2,U_3,\mu_2)
\]
along with $T_1, T_2,T_3$. We require that
$\AdvB_{\kappa_1,\kappa_2}$ decides whether $T_1, T_2, T_3$ are
distributed as
$g_2^{\tau_1\vecb_1^\ast},g_2^{\tau_1\vecb_2^\ast},g_2^{\tau_1\vecb_3^\ast}$
or
$g_2^{\tau_1\vecb_1^\ast+\tau_2\vecb_4^\ast},g_2^{\tau_1\vecb_2^\ast+\tau_2\vecb_5^\ast},g_2^{\tau_1\vecb_3^\ast+\tau_2\vecb_6^\ast}$.

$\AdvB_{\kappa_1,\kappa_2}$ simulates $\Gm_{\kappa_1,\kappa_2}$ or
$\Gm_{\kappa_1,\kappa_2-1}$ with $\AdvA$, depending on the
distribution of $T_1, T_2, T_3$. To compute the public parameters
and master key, $\AdvB_{\kappa_1,\kappa_2}$ chooses a random matrix
$\matA \in \Z_q^{3\times 3}$ (with all but negligible probability,
$\matA$ is invertible). We then implicitly set dual orthonormal
bases $\Ddual,\Ddual^\ast$ to:
\begin{align*}
&\vecd_1:=\vecb_1, \quad\vecd_2:=\vecb_2, \quad\vecd_3:=\vecb_3,\quad (\vecd_4,\vecd_5,\vecd_6):=(\vecb_4,\vecb_5,\vecb_6)\matA,\\
&\vecd_1^\ast:=\vecb_1^\ast, \quad\vecd_2^\ast:=\vecb_2^\ast, \quad\vecd_3^\ast:=\vecb_3^\ast,
\quad (\vecd_4^\ast,\vecd_5^\ast,\vecd_6^\ast):=(\vecb_4^\ast,\vecb_5^\ast,\vecb_6^\ast)(\matA^{-1})^\tran.
\end{align*}
\noindent We note that $\Ddual,\Ddual^\ast$ are properly
distributed, and reveal no information about $\matA$.
$\AdvB_{\kappa_1,\kappa_2}$ chooses random value $\alpha\in\Z_q$ and
compute $g_T^\alpha:=e(g_1, g_2)^{\alpha\vecd_1\cdot\vecd_1^\ast}$. $\AdvB$ can
give $\AdvA$ the public parameters
\[
\PP:=\{\mathbb{G}; g_T^\alpha,g_1^{\vecd_1},g_1^{\vecd_2},g_1^{\vecd_3}\}.
\]
\noindent The master key
\[
\MK:=\{\alpha,g_2^{\vecd_1^\ast},g_2^{\vecd_2^\ast},g_2^{\vecd_3^\ast}\}
\]
is known to $\AdvB_{\kappa_1,\kappa_2}$, which allows
$\AdvB_{\kappa_1,\kappa_2}$ to respond to all of $\AdvA$'s private
key and key update queries by calling the normal key generation
algorithm. Since $\AdvB_{\kappa_1,\kappa_2}$ also knows
$g_2^{\vecd_4^\ast}$, $g_2^{\vecd_5^\ast}$, and
$g_2^{\vecd_6^\ast}$, it can easily produce semi-functional keys. To
answer the key queries that $\AdvA$ makes,
$\AdvB_{\kappa_1,\kappa_2}$ runs the semi-functional private key and
key update generation algorithm to produce semi-functional keys and
gives these to $\AdvA$. To answer the $\kappa_2$-th component of the
$(\kappa_1-\qnop)$-th key update for $\tp_{\kappa_1-\qnop}'$,
$\AdvB_{\kappa_1,\kappa_2}$ responds with:
\[
\sKp_{\tp_{\kappa_1-\qnop}',\theta}:=(g_2^{\vecb_1^\ast})^{\alpha_{\theta,2}} T_1^{\tp_{\kappa_1-\qnop}'}(T_3)^{-1}.
\]
\noindent This implicitly sets $r_{\theta,2}:=\tau_1$. If $T_1,
T_2,T_3$ are equal to
$g_2^{\tau_1\vecb_1^\ast},g_2^{\tau_1\vecb_2^\ast},g_2^{\tau_1\vecb_3^\ast}$,
then this is a properly distributed normal key update. Otherwise, if $T_1,
T_2,T_3$ are equal to
$g_2^{\tau_1\vecb_1^\ast+\tau_2\vecb_4^\ast},g_2^{\tau_1\vecb_2^\ast+\tau_2\vecb_5^\ast},g_2^{\tau_1\vecb_3^\ast+\tau_2\vecb_6^\ast}$,
then this is a semi-functional key update, whose exponent vector
includes
\begin{align}\label{equ3-ribe}
\tp_{\kappa_1-\qnop}'\tau_2\vecb_4^\ast-\tau_2\vecb_6^\ast
\end{align}
as its component in the span of
$\vecb_4^\ast,\vecb_5^\ast,\vecb_6^\ast$. To respond to the
remaining key queries, $\AdvB_{\kappa_1,\kappa_2}$ simply runs the
normal key generation algorithm.

At some point, $\AdvA$ sends $\AdvB_{\kappa_1,\kappa_2}$
two pairs $(\idp_{(0)}^\ast,\tp_{(0)}^\ast,\msgp_{(0)}^\ast)$ and
$(\idp_{(1)}^\ast,\tp_{(1)}^\ast,\msgp_{(1)}^\ast)$. $\AdvB_0$
chooses a random bit $\beta \in\{0, 1\}$ and encrypts
$\msgp_{(\beta)}$ under $(\idp^\ast_{(\beta)},\tp^\ast_{(\beta)})$
as follows:
\[
\sCp:=\msgp^\ast_{(\beta)}\cdot\left( e(U_1,g_2^{\vecb_1^\ast})\right)^{\alpha}=\msgp^\ast_{(\beta)}\cdot (g_T^\alpha)^z,
\quad \sCp_0:=U_1(U_2)^{\idp^\ast_{(\beta)}}(U_3)^{\tp^\ast_{(\beta)}},
\]
\noindent where $\AdvB_{\kappa_1,\kappa_2}$ has implicitly set
$z:=\mu_1$. The ``semi-functional part'' of the exponent vector here
is:
\begin{align}\label{equ4-ribe}
\mu_2\vecb_4+\idp^\ast_{(\beta)}\mu_2\vecb_5+\tp^\ast_{(\beta)}\mu_2\vecb_6.
\end{align}
We observe that if $\tp^\ast_{(\beta)} = \tp_{\kappa_1-\qnop}'$
(which is impossible), then vectors~\ref{equ3-ribe} and~\ref{equ4-ribe} would
be orthogonal, resulting in a nominally semi-functional ciphertext
and key pair ($\AdvB_{\kappa_1,\kappa_2}$ can also use $T_1,
T_2,T_3$ to generate private key part for $\idp^\ast_{(\beta)}$) of
Type I. It gives the ciphertext
$(\sCp,\sCp_0)$ to $\AdvA$.

We now argue that since $\tp^\ast_{(\beta)}\neq
\tp_{\kappa_1-\qnop}'$, in $\AdvA$'s view the vectors~\ref{equ3-ribe}
and~\ref{equ4-ribe} are distributed as random vectors in the spans of
$\vecd_4^\ast,\vecd_5^\ast,\vecd_6^\ast$ and
$\vecd_4,\vecd_5,\vecd_6$ respectively. To see this, we take the
coefficients of vectors~\ref{equ3-ribe} and~\ref{equ4-ribe} in terms of the
bases $\vecb_4^\ast,\vecb_5^\ast,\vecb_6^\ast$ and
$\vecb_4,\vecb_5,\vecb_6$ respectively and translate them into
coefficients in terms of the bases
$\vecd_4^\ast,\vecd_5^\ast,\vecd_6^\ast$ and
$\vecd_4,\vecd_5,\vecd_6$. Using the change of basis matrix $\matA$,
we obtain the new coefficients (in vector form) as:
\[
\tau_2\matA^\tran(\tp_{\kappa_1-\qnop}',-1,0)^\tran,\, \mu_2\matA^{-1}(1,\idp^\ast_{(\beta)},\tp^\ast_{(\beta)})^\tran.
\]
Since the distribution of everything given to $\matA$ except for the
$\kappa_2$-th component of the $(\kappa_1-\qnop)$-th key update
$\sKp_{\tp_{\kappa_1-\qnop}',\theta}$ and the challenge ciphertext
$(\sCp,\sCp_0)$ is independent
of the random matrix $\matA$ and $\tp^\ast_{(\beta)}\neq
\tp_{\kappa_1-\qnop}'$,
we can conclude
that these coefficients are uniformly except with probability $4/q$ (namely, the cases $\mu_2$ or $\tau_2$ defined in Subspace problem is zero, $(\chi_4,\chi_5,\chi_6)$ or $(\nu_{\theta,4,2},\nu_{\theta,5,2},\nu_{\theta,6,2})$ defined in Equations~\ref{equ:semi-ct-ribe} and~\ref{equ:semi-key-ribe-2} is the zero vector) from Lemma~\ref{lem:mat-indist}. Thus, $\AdvB_{\kappa_1,\kappa_2}$ has
properly simulated $\Gm_{\kappa_1,\kappa_2}$ in this case.

If $T_1,T_2, T_3$ are equal to
$g_2^{\tau_1\vecb_1^*},g_2^{\tau_1\vecb_2^*},g_2^{\tau_1\vecb_3^*}$,
then the coefficients of the vector~\ref{equ4-ribe} are uniformly except with probability $2/q$ (namely, the cases $\mu_2$ defined in Subspace problem is zero, $(\chi_4,\chi_5,\chi_6)$ defined in Equations~\ref{equ:semi-ct-ribe} is the zero vector) from Lemma~\ref{lem:mat-indist}. Thus, $\AdvB_{\kappa_1,\kappa_2}$ has
properly simulated $\Gm_{\kappa_1,\kappa_2-1}$ in this case.

In summary, $\AdvB_{\kappa_1,\kappa_2}$ has properly simulated
either $\Gm_{\kappa_1,\kappa_2-1}$ or $\Gm_{\kappa_1,\kappa_2}$ for
$\AdvA$, depending on the distribution of $T_1, T_2,T_3$. It can
therefore leverage $\AdvA$'s advantage $\epsilon$ between these
games to obtain an advantage $\epsilon + \frac{6}{q}$ against
the Subspace assumption in $G_2$, namely
$\Adv_{\AdvB_\kappa}^{\DStp}(\lambda)=\epsilon +
\frac{6}{q}$.
\end{IEEEproof}

\begin{mylem} \label{lem:security-ribe-5}
For $\kappa_2$ from $0$ to $4N_{max}$, suppose that there exists an adversary $\AdvA$ where $|\Adv_{\AdvA}^{\Gm_{\qnop+\qntp+1,\kappa_2-1}}(\lambda)-\Adv_{\AdvA}^{\Gm_{\qnop+\qntp+1,\kappa_2}}(\lambda)|=\epsilon$.
Then there exists an algorithm $\AdvB_{\qnop+\qntp+1,\kappa_2}$ such that $\Adv_{\AdvB_{\qnop+\qntp+1,\kappa_2}}^{\DStp}(\lambda)=\epsilon+ \frac{8}{q}$, with $K=3$ and $N=6$.
\end{mylem}
\begin{IEEEproof}
$\AdvB_{\qnop+\qntp+1,\kappa_2}$ is given
\[
D:=(\mathbb{G};g_1^{\vecb_1},g_1^{\vecb_2},g_1^{\vecb_3},g_2^{\vecb_1^\ast},\ldots,g_2^{\vecb_6^\ast},U_1,U_2,U_3,\mu_2)
\]
along with $T_1, T_2,T_3$. We require that
$\AdvB_{\qnop+\qntp+1,\kappa_2}$ decides whether $T_1, T_2, T_3$ are
distributed as
$g_2^{\tau_1\vecb_1^\ast},g_2^{\tau_1\vecb_2^\ast},g_2^{\tau_1\vecb_3^\ast}$
or
$g_2^{\tau_1\vecb_1^\ast+\tau_2\vecb_4^\ast},g_2^{\tau_1\vecb_2^\ast+\tau_2\vecb_5^\ast},g_2^{\tau_1\vecb_3^\ast+\tau_2\vecb_6^\ast}$.

$\AdvB_{\qnop+\qntp+1,\kappa_2}$ simulates
$\Gm_{\qnop+\qntp+1,\kappa_2}$ or $\Gm_{\qnop+\qntp+1,\kappa_2-1}$
with $\AdvA$, depending on the distribution of $T_1, T_2, T_3$. To
compute the public parameters and master key,
$\AdvB_{\qnop+\qntp+1,\kappa_2}$ chooses a random matrix $\matA \in
\Z_q^{3\times 3}$ (with all but negligible probability, $\matA$ is
invertible). We then implicitly set dual orthonormal bases
$\Ddual,\Ddual^\ast$ to:
\begin{align*}
&\vecd_1:=\vecb_1, \quad\vecd_2:=\vecb_2, \quad\vecd_3:=\vecb_3,\quad (\vecd_4,\vecd_5,\vecd_6):=(\vecb_4,\vecb_5,\vecb_6)\matA,\\
&\vecd_1^\ast:=\vecb_1^\ast, \quad\vecd_2^\ast:=\vecb_2^\ast, \quad\vecd_3^\ast:=\vecb_3^\ast,
\quad (\vecd_4^\ast,\vecd_5^\ast,\vecd_6^\ast):=(\vecb_4^\ast,\vecb_5^\ast,\vecb_6^\ast)(\matA^{-1})^\tran.
\end{align*}
\noindent We note that $\Ddual,\Ddual^\ast$ are properly
distributed, and reveal no information about $\matA$.
$\AdvB_{\qnop+\qntp+1,\kappa_2}$ chooses random value
$\alpha\in\Z_q$ and compute $g_T^\alpha:=e(g_1,
g_2)^{\alpha\vecd_1\cdot\vecd_1^\ast}$. $\AdvB$ can gives $\AdvA$
the public parameters
\[
\PP:=\{\mathbb{G}; g_T^\alpha,g_1^{\vecd_1},g_1^{\vecd_2},g_1^{\vecd_3}\}.
\]
\noindent The master key
\[
\MK:=\{\alpha,g_2^{\vecd_1^\ast},g_2^{\vecd_2^\ast},g_2^{\vecd_3^\ast}\}
\]
is known to $\AdvB_{\qnop+\qntp+1,\kappa_2}$, which allows
$\AdvB_{\qnop+\qntp+1,\kappa_2}$ to respond to all of $\AdvA$'s
private key and key update queries by calling the normal key
generation algorithm. Since $\AdvB_{\qnop+\qntp+1,\kappa_2}$ also
knows $g_2^{\vecd_4^\ast}$, $g_2^{\vecd_5^\ast}$, and
$g_2^{\vecd_6^\ast}$, it can easily produce semi-functional keys. To
answer the key queries that $\AdvA$ makes,
$\AdvB_{\qnop+\qntp+1,\kappa_2}$ runs the semi-functional private
key and key update generation algorithm to produce semi-functional
keys and gives these to $\AdvA$.

However, $\AdvB_{\qnop+\qntp+1,\kappa_2}$ changes the strategy to
respond all the components for the $\kappa_2$-th node in the binary
tree of private keys and key updates. All key queries for $\Gamma_1$
and $\Gamma_2$ are similar with the following process except that
$\AdvB_{\qnop+\qntp+1,\kappa_2}$ uses
$g_2^{\vecd_1^\ast},\ldots,g_2^{\vecd_6^\ast}$ to re-randomize the
exponents. To answer the component for the challenge identities
$\idp_{(0)}^\ast,\idp_{(1)}^\ast$ and times
$\tp_{(0)}^\ast,\tp_{(1)}^\ast$ (namely, the $\phi_1,\phi_2$-th
private key and $\phi_3,\phi_4$-th key update queries) on the
$\kappa_2$-th node, $\AdvB_{\qnop+\qntp+1,\kappa_2}$ picks
$\alpha_{\theta,1}',\alpha_{\theta,1}''\in\Z_q$ and responds with:
\begin{align*}
&\sKp_{\idp_{(0)}^\ast,\theta}:=g_2^{\alpha_{\theta,1}'\vecb_1^\ast}(T_1^{\vecb_1^\ast})^{\alpha_{\theta,1}''} T_1^{r_{\theta,1}'\idp_{(0)}^\ast}(T_2)^{-r_{\theta,1}'},\\
&\sKp_{\tp_{(0)}^\ast,\theta}:=g_2^{(\alpha-\alpha_{\theta,1}')}(T_1^{\vecb_1^\ast})^{-\alpha_{\theta,1}''} T_1^{r_{\theta,2}'\tp_{(0)}^\ast}(T_2)^{-r_{\theta,2}'},\\
&\sKp_{\idp_{(1)}^\ast,\theta}:=g_2^{\alpha_{\theta,1}'\vecb_1^\ast}(T_1^{\vecb_1^\ast})^{\alpha_{\theta,1}''} T_1^{r_{\theta,1}''\idp_{(1)}^\ast}(T_2)^{-r_{\theta,1}''},\\
&\sKp_{\tp_{(1)}^\ast,\theta}:=g_2^{(\alpha-\alpha_{\theta,1}')}(T_1^{\vecb_1^\ast})^{-\alpha_{\theta,1}''} T_1^{r_{\theta,2}''\tp_{(1)}^\ast}(T_2)^{-r_{\theta,2}''},
\end{align*}
where $\AdvB_{\qnop+\qntp+1,\kappa_2}$ implicitly sets
$\alpha_{\theta,1}:=\alpha_{\theta,1}'+\alpha_{\theta,1}''\tau_1$
and
$\alpha_{\theta,2}:=\alpha-\alpha_{\theta,1}'-\alpha_{\theta,1}''\tau_1$
(note that $\alpha_{\theta,1}+\alpha_{\theta,2}=\alpha$). Note that
from the restriction of queries for the challenge identities and
times, only part of the keys are given to $\AdvA$.

If $T_1, T_2, T_3$ are equal to
$g_2^{\tau_1\vecb_1^\ast},g_2^{\tau_1\vecb_2^\ast},g_2^{\tau_1\vecb_3^\ast}$,
then these are properly distributed normal keys. If $T_1, T_2, T_3$
are equal to
$g_2^{\tau_1\vecb_1^\ast+\tau_2\vecb_4^\ast},g_2^{\tau_1\vecb_2^\ast+\tau_2\vecb_5^\ast},g_2^{\tau_1\vecb_3^\ast+\tau_2\vecb_6^\ast}$,
then these are semi-functional keys, whose exponent vector includes
\begin{align}
(\alpha_{\theta,1}''\tau_2 + \idp_{(0)}^\ast \tau_2 r_{\theta,1}')\vecb_4^\ast-\tau_2r_{\theta,1}'\vecb_5^\ast,\label{equ5-ribe}\\
(-\alpha_{\theta,1}''\tau_2 + \tp_{(0)}^\ast \tau_2 r_{\theta,2}')\vecb_4^\ast-\tau_2r_{\theta,2}'\vecb_6^\ast,\label{equ6-ribe}\\
(\alpha_{\theta,1}''\tau_2 + \idp_{(1)}^\ast \tau_2 r_{\theta,1}'')\vecb_4^\ast-\tau_2r_{\theta,1}''\vecb_5^\ast,\label{equ7-ribe}\\
(-\alpha_{\theta,1}''\tau_2 + \tp_{(1)}^\ast \tau_2 r_{\theta,2}'')\vecb_4^\ast-\tau_2r_{\theta,2}''\vecb_6^\ast,\label{equ8-ribe}
\end{align}
as its component in the span of
$\vecb_4^\ast,\vecb_5^\ast,\vecb_6^\ast$ respectively. To respond to
the remaining key queries, $\AdvB_{\qnop+\qntp+1,\kappa_2}$ simply
runs the normal key generation algorithm.

At some point, $\AdvA$ sends $\AdvB_{\qnop+\qntp+1,\kappa_2}$ two challenge pairs $(\idp_{(0)}^\ast,\tp_{(0)}^\ast,\msgp_{(0)}^\ast)$
and $(\idp_{(1)}^\ast,\tp_{(1)}^\ast,\msgp_{(1)}^\ast)$. $\AdvB_{\qnop+\qntp+1,\kappa_2}$
chooses a random bit $\beta \in\{0, 1\}$ and encrypts
$\msgp^\ast_{(\beta)}$ under $(\idp^\ast_{(\beta)},\tp^\ast_{(\beta)})$
as follows:
\[
\sCp:=\msgp^\ast_{(\beta)}\cdot\left( e(U_1,g_2^{\vecb_1^\ast})\right)^{\alpha}=\msgp^\ast_{(\beta)}\cdot (g_T^\alpha)^z,
\quad \sCp_0:=U_1(U_2)^{\idp^\ast_{(\beta)}}(U_3)^{\tp^\ast_{(\beta)}},
\]
\noindent where $\AdvB_{\qnop+\qntp+1,\kappa_2}$ has implicitly set
$z:=\mu_1$. The ``semi-functional part'' of the exponent vector here
is:
\begin{align}\label{equ9-ribe}
\mu_2\vecb_4+\idp^\ast_{(\beta)}\mu_2\vecb_5+\tp^\ast_{(\beta)}\mu_2\vecb_6.
\end{align}
We observe that
$((\sCp,\sCp_0),\sKp_{\idp_{(\beta)}^\ast,\theta},\sKp_{\tp_{(\beta)}^\ast,\theta})$
would result in a nominally semi-functional ciphertext and key pair
of Type II. It gives the ciphertext
$(\sCp,\sCp_0)$ to $\AdvA$.

Since the adversary $\AdvA$ is only allowed to query one of the
following sets for the challenge identities and times:
\[
\emptyset, ~\{\idp_{(0)}^\ast\},~\{\idp_{(1)}^\ast\}, ~\{\tp_{(0)}^\ast\},~\{\tp_{(1)}^\ast\}, ~\{\idp_{(0)}^\ast,\idp_{(1)}^\ast\}, ~\{\tp_{(0)}^\ast,\tp_{(1)}^\ast\},
\]
\[
\{\idp_{(0)}^\ast,\tp_{(1)}^\ast\},~\{\idp_{(1)}^\ast,\tp_{(0)}^\ast\},
\]
we now argue that in $\AdvA$'s view the given
vectors~\ref{equ5-ribe},~\ref{equ6-ribe},~\ref{equ7-ribe},~\ref{equ8-ribe} and \ref{equ9-ribe} are distributed as
random vectors in the spans of
$\vecd_4^\ast,\vecd_5^\ast,\vecd_6^\ast$ and
$\vecd_4,\vecd_5,\vecd_6$ respectively. To see this, we take the
coefficients of vectors~\ref{equ5-ribe},~\ref{equ6-ribe},~\ref{equ7-ribe},~\ref{equ8-ribe} and \ref{equ9-ribe} in
terms of the bases $\vecb_4^\ast,\vecb_5^\ast,\vecb_6^\ast$ and
$\vecb_4,\vecb_5,\vecb_6$ respectively and translate them into
coefficients in terms of the bases
$\vecd_4^\ast,\vecd_5^\ast,\vecd_6^\ast$ and
$\vecd_4,\vecd_5,\vecd_6$. Using the change of basis matrix $\matA$
and statistical indistinguishability lemmas, we obtain new random
coefficients (in vector form), which are summarized in the following Table:
\begin{table}[H]
\centering
\begin{tabular}{|c|c|c|c|}
\hline
Case & Type of Adversary &  New Coefficients\\
\hline
1& $\emptyset$ & $\mu_2\matA^{-1}(1,\idp^\ast_{(\beta)},\tp^\ast_{(\beta)})^\tran$ \\
\hline
2&$\{\idp_{(0)}^\ast\}$   & \makecell{$\mu_2\matA^{-1}(1,\idp^\ast_{(\beta)},\tp^\ast_{(\beta)})^\tran$\\ $  \matA^\tran(\alpha_{\theta,1}''\tau_2 + \idp_{(0)}^\ast \tau_2r_{\theta,1}' ,- \tau_2 r_{\theta,1}',0)^\tran$ } \\
\hline
3&$\{\idp_{(1)}^\ast\}$   &  \makecell{$\mu_2\matA^{-1}(1,\idp^\ast_{(\beta)},\tp^\ast_{(\beta)})^\tran$\\ $\matA^\tran(\alpha_{\theta,1}''\tau_2 + \idp_{(1)}^\ast\tau_2 r_{\theta,1}'' ,-\tau_2 r_{\theta,1}'',0)^\tran$ } \\
\hline
4&$\{\tp_{(0)}^\ast\}$    & \makecell{$\mu_2\matA^{-1}(1,\idp^\ast_{(\beta)},\tp^\ast_{(\beta)})^\tran$\\ $\matA^\tran(-\alpha_{\theta,1}''\tau_2 + \tp_{(0)}^\ast \tau_2 r_{\theta,2}',0,-\tau_2r_{\theta,2}')^\tran$} \\
\hline
5&$\{\tp_{(1)}^\ast\}$    & \makecell{$\mu_2\matA^{-1}(1,\idp^\ast_{(\beta)},\tp^\ast_{(\beta)})^\tran$\\ $\matA^\tran(-\alpha_{\theta,1}''\tau_2 + \tp_{(1)}^\ast \tau_2 r_{\theta,2}'',0,-\tau_2 r_{\theta,2}'')^\tran$ }  \\
\hline
6&$\{\idp_{(0)}^\ast,\idp_{(1)}^\ast\}$  & \makecell{$\mu_2\matA^{-1}(1,\idp^\ast_{(\beta)},\tp^\ast_{(\beta)})^\tran$\\ $ \matA^\tran(\alpha_{\theta,1}''\tau_2 + \idp_{(0)}^\ast \tau_2 r_{\theta,1}' ,- \tau_2 r_{\theta,1}',0)^\tran$ \\ $\matA^\tran(\alpha_{\theta,1}''\tau_2 + \idp_{(1)}^\ast \tau_2 r_{\theta,1}'',-\tau_2 r_{\theta,1}'',0)^\tran$ } \\
\hline
7&$\{\tp_{(0)}^\ast,\tp_{(1)}^\ast\}$  & \makecell{$\mu_2\matA^{-1}(1,\idp^\ast_{(\beta)},\tp^\ast_{(\beta)})^\tran$\\ $ \matA^\tran(-\alpha_{\theta,1}''\tau_2 + \tp_{(0)}^\ast \tau_2 r_{\theta,2}',0,-\tau_2 r_{\theta,2}')^\tran$ \\ $\matA^\tran(-\alpha_{\theta,1}''\tau_2 + \tp_{(1)}^\ast \tau_2 r_{\theta,2}'',0,-\tau_2 r_{\theta,2}'')^\tran$ } \\
\hline
8&$\{\idp_{(0)}^\ast,\tp_{(1)}^\ast\}$  & \makecell{$\mu_2\matA^{-1}(1,\idp^\ast_{(\beta)},\tp^\ast_{(\beta)})^\tran$\\ $ \matA^\tran(\alpha_{\theta,1}''\tau_2 + \idp_{(0)}^\ast \tau_2 r_{\theta,1}' ,- \tau_2 r_{\theta,1}',0)^\tran$ \\ $\matA^\tran(-\alpha_{\theta,1}''\tau_2 + \tp_{(1)}^\ast\tau_2 r_{\theta,2}'' ,0,-\tau_2 r_{\theta,2}'')^\tran$ } \\
\hline
9&$\{\idp_{(1)}^\ast,\tp_{(0)}^\ast\}$  &  \makecell{$\mu_2\matA^{-1}(1,\idp^\ast_{(\beta)},\tp^\ast_{(\beta)})^\tran$\\ $\matA^\tran(\alpha_{\theta,1}''\tau_2 + \idp_{(1)}^\ast \tau_2 r_{\theta,1}'',-\tau_2 r_{\theta,1}'',0)^\tran$ \\$ \matA^\tran(-\alpha_{\theta,1}''\tau_2 + \tp_{(0)}^\ast \tau_2 r_{\theta,2}',0,-\tau_2 r_{\theta,2}')^\tran$ } \\
\hline
\end{tabular}\label{tab:adversary}
\end{table}
\noindent Since the distribution of everything given to $\matA$
except for the coefficients of the vectors in above Table is independent of the random matrix $\matA$,
we can conclude
that these coefficients are uniformly except with probability
\begin{itemize}
  \item  $2/q$, namely except for the cases:
    \begin{itemize}
    \item  $\mu_2$ defined in Subspace problem is zero,
    \item  $(\chi_4,\chi_5,\chi_6)$ defined in Equation~\ref{equ:semi-ct-ribe} the zero vector,
    \end{itemize}
    from Lemma~\ref{lem:mat-indist} for Case 1.

  \item $4/q$, namely except for the cases:
    \begin{itemize}
    \item  $\mu_2$ or $\tau_2$ defined in Subspace problem is zero,
    \item  $(\chi_4,\chi_5,\chi_6)$ or $(\nu_{\theta,4,1},\nu_{\theta,5,1},\nu_{\theta,6,1})$ or $(\nu_{\theta,4,2},\nu_{\theta,5,2},\nu_{\theta,6,2})$ defined in Equations~\ref{equ:semi-ct-ribe},~\ref{equ:semi-key-ribe-1} and~\ref{equ:semi-key-ribe-2} is the zero vector,
    \end{itemize}
    from Lemma~\ref{lem:mat-indist1} for Cases 2-5, since $\alpha_{\theta,1}''$ is randomly picked from $\Z_q$.

  \item $6/q$, namely except for the cases:
    \begin{itemize}
    \item  $\mu_2$ or $\tau_2$ defined in Subspace problem is zero,
    \item  $(\chi_4,\chi_5,\chi_6)$ or $(\nu_{\theta,4,1},\nu_{\theta,5,1},\nu_{\theta,6,1})$ or $(\nu_{\theta,4,2},\nu_{\theta,5,2},\nu_{\theta,6,2})$ defined in Equations~\ref{equ:semi-ct-ribe},~\ref{equ:semi-key-ribe-1} and~\ref{equ:semi-key-ribe-2} is the zero vector,
    \item $(\nu_{\theta,4,1},\nu_{\theta,5,1},\nu_{\theta,6,1})$ and $(\nu_{\theta,4,2},\nu_{\theta,5,2},\nu_{\theta,6,2})$ defined in Equations~\ref{equ:semi-key-ribe-1} and \ref{equ:semi-key-ribe-2} are linearly dependent,
    \end{itemize}
    from Lemma~\ref{lem:mat-indist} for Cases 6-9, since $\alpha_{\theta,1}''$ is randomly picked from $\Z_q$ and the
        coefficients of vectors~\ref{equ5-ribe},~\ref{equ6-ribe},~\ref{equ7-ribe},~\ref{equ8-ribe} are linearly independent.

\end{itemize}
Thus, $\AdvB_{\qnop+\qntp+1,\kappa_2}$ has
properly simulated $\Gm_{\qnop+\qntp+1,\kappa_2}$  in this case.

If $T_1,T_2, T_3$ are equal to
$g_2^{\tau_1\vecb_1^*},g_2^{\tau_1\vecb_2^*},g_2^{\tau_1\vecb_3^*}$,
then the coefficients of the vector~\ref{equ9-ribe} are uniformly except with probability $2/q$ (namely, the cases $\mu_2$ defined in Subspace problem is zero, $(\chi_4,\chi_5,\chi_6)$ defined in Equations~\ref{equ:semi-ct-ribe} is the zero vector) from Lemma~\ref{lem:mat-indist}. Thus, $\AdvB_{\qnop+\qntp+1,\kappa_2}$ has
properly simulated $\Gm_{\qnop+\qntp+1,\kappa_2-1}$ in this case.

In summary, $\AdvB_{\qnop+\qntp+1,\kappa_2}$ has properly simulated
either $\Gm_{\qnop+\qntp+1,\kappa_2-1}$ or $\Gm_{\qnop+\qntp+1,\kappa_2}$ for
$\AdvA$, depending on the distribution of $T_1, T_2,T_3$. It can
therefore leverage $\AdvA$'s advantage $\epsilon$ between these
games to obtain an advantage $\epsilon + \frac{8}{q}$ against
the Subspace assumption in $G_2$, namely
$\Adv_{\AdvB_{\qnop+\qntp+1,\kappa_2}}^{\DStp}(\lambda)=\epsilon+
\frac{8}{q}$.
\end{IEEEproof}

\begin{mylem} \label{lem:security-ribe-6}
For any adversary $\AdvA$, $\Adv_{\AdvA}^{\Gm_{\qnop+\qntp+1,4N_{max}}}(\lambda)\leq\Adv_{\AdvA}^{\Gm_{Final}}(\lambda)+\frac{1}{q}$.
\end{mylem}
\begin{IEEEproof}
To prove this lemma, we show the joint distributions of
\[(\PP,\CT_{\idp^\ast_{(\beta)},\tp^\ast_{(\beta)}}^{(\SFp)},\{\SK_{\idp_\ell}^{(\SFp)}\}_{\ell\in[\qnop]},\{\KU_{\tp_\ell}^{(\SFp)}\}_{\ell\in[\qntp]})\]
in $\Gm_{\nu}$ and that of
\[(\PP,\CT_{\idp_{(\Rp)},\tp_{(\Rp)}}^{(\Rp)},\{\SK_{\idp_\ell}^{(\SFp)}\}_{\ell\in[\qnop]},\{\KU_{\tp_\ell}^{(\SFp)}\}_{\ell\in[\qntp]})\]
in $\Gm_{Final}$ are equivalent for the adversary's view, where
$\CT_{\idp_{(\Rp)},\tp_{(\Rp)}}^{(\Rp)}$ is a semi-functional encryption of a
random message in $G_T$ and under a random identity $\idp_{(\Rp)}$ in $\Z_q$ and a
random time $\tp_{(\Rp)}$ in $\Z_q$.

For this purpose, we pick $\matA:=(\xi_{i,j})\getsr \Z_q^{3\times 3}$
and define new dual orthonormal bases
$\Fdual:=(\vecf_1,\ldots,\vecf_6)$, and
$\Fdual^\ast:=(\vecf_1^\ast,\ldots,\vecf_6^\ast)$ as follows:
\begin{align*}
&\left(
  \begin{array}{c}
    \vecf_1\\
    \vecf_2\\
    \vecf_3\\
    \vecf_4\\
    \vecf_5\\
    \vecf_6
  \end{array}
\right)
:=
\left(
  \begin{array}{cccccc}
    1&0&0&0&0&0\\
    0&1&0&0&0&0\\
    0&0&1&0&0&0\\
    \xi_{1,1}&\xi_{1,2}&\xi_{1,3}&1&0&0\\
    \xi_{2,1}&\xi_{2,2}&\xi_{2,3}&0&1&0\\
    \xi_{3,1}&\xi_{3,2}&\xi_{3,3}&0&0&1
  \end{array}
\right)
\left(
  \begin{array}{c}
    \vecd_1\\
    \vecd_2\\
    \vecd_3\\
    \vecd_4\\
    \vecd_5\\
    \vecd_6
  \end{array}
\right),\\
&\left(
  \begin{array}{c}
    \vecf_1^\ast\\
    \vecf_2^\ast\\
    \vecf_3^\ast\\
    \vecf_4^\ast\\
    \vecf_5^\ast\\
    \vecf_6^\ast
  \end{array}
\right)
:=
\left(
  \begin{array}{cccccc}
    1&0&0&-\xi_{1,1}&-\xi_{2,1}&-\xi_{3,1}\\
    0&1&0&-\xi_{1,2}&-\xi_{2,2}&-\xi_{3,2}\\
    0&0&1&-\xi_{1,3}&-\xi_{2,3}&-\xi_{3,3}\\
    0&0&0&1&0&0\\
    0&0&0&0&1&0\\
    0&0&0&0&0&1
  \end{array}
\right)
\left(
  \begin{array}{c}
    \vecd_1^\ast\\
    \vecd_2^\ast\\
    \vecd_3^\ast\\
    \vecd_4^\ast\\
    \vecd_5^\ast\\
    \vecd_6^\ast
  \end{array}
\right).
\end{align*}
It is easy to verify that $\Fdual$ and $\Fdual^\ast$ are also dual
orthonormal, and are distributed the same as $\Ddual$ and
$\Ddual^\ast$.

Then the public parameters, challenge ciphertext, queried
private keys and key updates
in $\Gm_{\qnop+\qntp+1,4N_{max}}$ are expressed over bases $\Ddual$ and
$\Ddual^\ast$ as
\begin{align*}
&\PP:=\{\mathbb{G}; g_T^\alpha,g_1^{\vecd_1},g_1^{\vecd_2},g_1^{\vecd_3}\},\\
&\CT_{\idp^\ast_{(\beta)}}^{(\SFp)}:=\biggl\{\sCp:=\msgp\cdot (g_T^\alpha)^z, \quad \sCp_0:=g_1^{z( \vecd_1 + \idp \vecd_2 +\tp\vecd_3 ) + \chi_4\vecd_4 + \chi_5\vecd_5 + \chi_6\vecd_6} \biggr\},\\
&\biggl\{
\SK_{\idp_\ell}^{(\SFp)}:=\left\{
\left(\theta,  \sKp_{\idp_\ell,\theta}^{(\SFp)}:=g_2^{( \alpha_{\theta,1} + r_{\theta,1} \idp_\ell ) \vecd_1^\ast - r_{\theta,1} \vecd_2^\ast + \nu_{\theta,4,1}\vecd_4^\ast + \nu_{\theta,5,1}\vecd_5^\ast + \nu_{\theta,6,1}\vecd_6^\ast} \right)\right\}_{\theta\in \Path(v_\ell)} \biggr\}_{\ell\in[\qnop]},\\
&\biggl\{
\KU_{\tp_\ell}^{(\SFp)}:=\left\{
\left(\theta, \sKp_{\tp_\ell,\theta}^{(\SFp)}:=g_2^{( \alpha_{\theta,2} + r_{\theta,2} \tp_\ell ) \vecd_1^\ast - r_{\theta,2} \vecd_3^\ast  + \nu_{\theta,4,2}\vecd_4^\ast + \nu_{\theta,5,2}\vecd_5^\ast + \nu_{\theta,6,2}\vecd_6^\ast} \right)\right\}_{\theta\in\KUNodes(\BT, \RLp, \tp_\ell)}\biggr\}_{\ell\in[\qntp]}.
\end{align*}

\noindent Then we can express them over bases $\Fdual$ and $\Fdual^\ast$ as
\begin{align*}
&\PP:=\{\mathbb{G}; g_T^\alpha,g_1^{\vecf_1},g_1^{\vecf_2},g_1^{\vecf_3}\},\\
&\CT_{\idp^\ast_{(\beta)}}^{(\SFp)}:=\biggl\{\sCp:=\msgp\cdot (g_T^\alpha)^z,\quad \sCp_0:=g_1^{z_1'\vecf_1 + z_2'\vecf_2 +z_3'\vecf_3 + \chi_4\vecf_4 + \chi_5\vecf_5 + \chi_6\vecf_6}\biggr\},\\
&\biggl\{
\SK_{\idp_\ell}^{(\SFp)}:=\left\{
\left(\theta, \sKp_{\idp_\ell,\theta}^{(\SFp)}:=g_2^{( \alpha_{\theta,1} + r_{\theta,1} \idp_\ell ) \vecf_1^\ast - r_{\theta,1} \vecf_2^\ast + \nu_{\theta,4,1}'\vecf_4^\ast + \nu_{\theta,5,1}'\vecf_5^\ast + \nu_{\theta,6,1}'\vecf_6^\ast} \right)\right\}_{\theta\in \Path(v_\ell)} \biggr\}_{\ell\in[\qnop]},\\
&\biggl\{
\KU_{\tp_\ell}^{(\SFp)}:=\left\{
\left(\theta, \sKp_{\tp_\ell,\theta}^{(\SFp)}:=g_2^{( \alpha_{\theta,2} + r_{\theta,2} \tp_\ell ) \vecf_1^\ast - r_{\theta,2} \vecf_3^\ast  + \nu_{\theta,4,2}'\vecf_4^\ast + \nu_{\theta,5,2}'\vecf_5^\ast +\nu_{\theta,6,2}'\vecf_6^\ast} \right)\right\}_{\theta\in\KUNodes(\BT, \RLp, \tp_\ell)} \biggr\}_{\ell\in[\qntp]}.
\end{align*}
where
\begin{align*}
&z_1':=z-\chi_4\xi_{1,1}-\chi_5\xi_{2,1}-\chi_6\xi_{3,1}, \\
&z_2':=z\idp^\ast_{(\beta)}-\chi_4\xi_{1,2}-\chi_5\xi_{2,2}-\chi_6\xi_{3,2}, \\
&z_3':=z\tp^\ast_{(\beta)}-\chi_4\xi_{1,3}-\chi_5\xi_{2,3}-\chi_6\xi_{3,3}\\
\end{align*}
\begin{align*}
&\left\{
\begin{array}{ll}
\nu_{\theta,4,1}':= \nu_{\theta,4,1}+\xi_{1,1}(\alpha_{\theta,1} + r_{\theta,1} \idp_\ell)-r_{\theta,1}\xi_{1,2}, \\
\nu_{\theta,5,1}':= \nu_{\theta,5,1}+\xi_{2,1}(\alpha_{\theta,1} + r_{\theta,1} \idp_\ell)-r_{\theta,1}\xi_{2,2}, \\
\nu_{\theta,6,1}':= \nu_{\theta,6,1}+\xi_{3,1}(\alpha_{\theta,1} + r_{\theta,1} \idp_\ell)-r_{\theta,1}\xi_{3,2}
\end{array}
\right\}
\end{align*}
for $\theta\in \Path(v_\ell), {\ell\in[\qnop]}$,\\
\begin{align*}
&\left\{
\begin{array}{ll}
\nu_{\theta,4,2}':= \nu_{\theta,4,2}+\xi_{1,1}(\alpha_{\theta,2} + r_{\theta,2} \tp_\ell)-r_{\theta,2}\xi_{1,3}, \\
\nu_{\theta,5,2}':= \nu_{\theta,5,2}+\xi_{2,1}(\alpha_{\theta,2} + r_{\theta,2} \tp_\ell)-r_{\theta,2}\xi_{2,3}, \\
\nu_{\theta,6,2}':= \nu_{\theta,6,2}+\xi_{3,1}(\alpha_{\theta,2} + r_{\theta,2} \tp_\ell)-r_{\theta,2}\xi_{3,3}
\end{array}
\right\}
\end{align*}
for ${\theta\in\KUNodes(\BT, \RLp, \tp_\ell)},{\ell\in[\qntp]} $, which are all uniformly distributed if $(\chi_4,\chi_5,\chi_6)$ defined in Equation~\ref{equ:semi-ct-ribe} is a non-zero vector since
\begin{align*}
&\{\xi_{ij}\}_{i\in[3],j\in[3]},\\
&\left\{\{\nu_{\theta,4,1},\nu_{\theta,5,1},\nu_{\theta,6,1}\}_{\theta\in
\Path(v_\ell)}\right\}_{\ell\in[\qnop]},\\
&\left\{\{\nu_{\theta,4,2}\nu_{\theta,5,2},
\nu_{\theta,6,2}\right\}_{\theta\in\KUNodes(\BT, \RLp,
\tp_\ell)}\}_{\ell\in[\qntp]}
\end{align*}
are all uniformly picked from
$\Z_q$.

In other words, the coefficients
$(z,z\idp^\ast_{(\beta)},z\tp^\ast_{(\beta)})$ of
$\vecd_1,\vecd_2,\vecd_3$ in the $\sCp_0$ term of the challenge
ciphertext is changed to random coefficients $(z_1',z_2',z_3')\in
\Z_q\times\Z_q\times\Z_q$ of $\vecf_1,\vecf_2,\vecf_3$, thus the
challenge ciphertext can be viewed as a semi-functional encryption
of a random message in $G_T$ and under a random identity in $\Z_q$
and a random time in $\Z_q$. Moreover, it is not difficult to check
that all other coefficients are well distributed. Thus
\[(\PP,\CT_{\idp^\ast_{(\beta)},\tp^\ast_{(\beta)}}^{(\SFp)},\{\SK_{\idp_\ell}^{(\SFp)}\}_{\ell\in[\qnop]},\{\KU_{\tp_\ell}^{(\SFp)}\}_{\ell\in[\qntp]})\]
expressed over bases $\Fdual$ and $\Fdual^\ast$ is properly
distributed as
\[(\PP,\CT_{\idp_{(\Rp)},\tp_{(\Rp)}}^{(\Rp)},\{\SK_{\idp_\ell}^{(\SFp)}\}_{\ell\in[\qnop]},\{\KU_{\tp_\ell}^{(\SFp)}\}_{\ell\in[\qntp]})\]
in $\Gm_{Final}$.

In the adversary's view, both $(\Ddual,\Ddual^\ast)$ and
$(\Fdual,\Fdual^\ast)$ are consistent with the same public key.
Therefore, the challenge ciphertext and queried secret keys above
can be expressed as keys and ciphertext in two ways, in $\Gm_{\nu}$
over bases $(\Ddual,\Ddual^\ast)$ and in $\Gm_{Final}$ over bases
$(\Fdual,\Fdual^\ast)$. Thus, $\Gm_{\qnop+\qnop+1,4N_{max}}$ and $\Gm_{Final}$ are
statistically indistinguishable.
\end{IEEEproof}

\begin{mylem}\label{lem:security-ribe-7}
For any adversary $\AdvA$, $\Adv_{\AdvA}^{\Gm_{Final}}(\lambda)=0$.
\end{mylem}
\begin{IEEEproof}
The value of $\beta$ is independent from the adversary's view in
$\Gm_{Final}$. Hence, $\Adv_{\AdvA}^{\Gm_{Final}}(\lambda)=0$.
\end{IEEEproof}


In $\Gm_{Final}$, the challenge ciphertext is a semi-functional
encryption of a random message in $G_T$ and under a random identity
in $\Z_q$ and a random time in $\Z_q$, independent of the two
messages, the challenge identities, and times provided by $\AdvA$.
Thus, our RIBE scheme is adaptively secure and anonymous.
\end{IEEEproof}

\section{Construction from DLIN}\label{dlin_construction}
We use the same binary tree structure mentioned in previous section in our second construction.
\subsection{Our Scheme}
Here we provide our second construction of RIBE under the DLIN assumption. Our RIBE scheme is specified as follows:

\begin{IEEEitemize}
\item $\Setup(\lambda,N_{max})$ On input a security parameter $\lambda$, a maximal number $N_{max}$ of users and generate a symmetric
bilinear pairing $\mathbb{G}:=(q,G,G_T,g,e)$ for sufficiently large
prime order $q$. Next perform the following steps:
\begin{IEEEenumerate}
\item Let $\RLp$ be an empty set and $\BT$ be a binary-tree with at least $N_{max}$ leaf nodes, set $\STp=\BT$.

\item Sample random dual orthonormal bases, $(\Ddual,\Ddual^\ast)\getsr \Dual(\Z_q^\dldimt)$. Let $\vecd_1,\ldots,\vecd_{\dldimt}$ denote the elements of $\Ddual$ and $\vecd_1^\ast,\ldots,\vecd_{\dldimt}^\ast$ denote the elements of $\Ddual^\ast$. It also picks $\alpha\getsr \Z_q$ and computes $g_T^\alpha:=e(g,g)^{\alpha\vecd_1\cdot \vecd_1^\ast}$.

\item Output $\RLp$, $\STp$, the public parameters
\begin{align*}
\PP:=\left\{\mathbb{G}; g_T^\alpha,g^{\vecd_1},\ldots,g^{\vecd_6}\right\},
\end{align*}
and the master key $\MK$
\[
\MK:=\left\{\alpha,g^{\vecd_1^\ast},\ldots,g^{\vecd_6^\ast}\right\}.
\]
\end{IEEEenumerate}

\item $\PriKeyGen(\PP,\MK,\idp,\RLp,\STp)$  On input the public parameters $\PP$, the master key $\MK$, an identity $\idp$, the revocation list $\RLp$, and the state $\STp$, it picks an unassigned leaf node $v$ from $\BT$ and stores $\idp$ in that node. It then performs the following steps:
\begin{IEEEenumerate}
    \item For any $\theta\in \Path(v)$, if $\alpha_{\theta,1},\alpha_{\theta,2},\alpha_{\theta,3}$ are undefined, then pick $\alpha_{\theta,1},\alpha_{\theta,3}$ $\getsr \Z_q$, set $\alpha_{\theta,2}=\alpha - \alpha_{\theta,1}$, and store them in node $\theta$. Pick $r_{\theta,1},r_{\theta,3}\getsr \Z_q$ and compute
        \[
        \sKp_{\idp,\theta}:=g^{ ( \alpha_{\theta,1} + r_{\theta,1} \idp ) \vecd_1^\ast - r_{\theta,1} \vecd_2^\ast + ( \alpha_{\theta,3} + r_{\theta,3} \idp ) \vecd_4^\ast - r_{\theta,3} \vecd_5^\ast }.
        \]

    \item Output $\SK_\idp:=\{(\theta,\sKp_{\idp,\theta})\}_{\theta\in \Path(v)}$, $\STp$.
\end{IEEEenumerate}

\item $\KeyUpd(\PP,\MK,\tp,\RLp,\STp)$ On input the public parameters $\PP$, the master key $\MK$, a time $\tp$, the revocation list $\RLp$, and the state $\STp$, it performs the following steps:
\begin{IEEEenumerate}
\item $\forall \theta\in\KUNodes(\BT, \RLp, \tp)$, if $\alpha_{\theta,1},\alpha_{\theta,2},\alpha_{\theta,3}$ are undefined, then pick $\alpha_{\theta,1},\alpha_{\theta,3}\getsr \Z_q$, set $\alpha_{\theta,2}=\alpha - \alpha_{\theta,1}$, and store them in node $\theta$. Pick $r_{\theta,2},r_{\theta,4}\getsr \Z_q$ and compute
        \[
        \sKp_{\tp,\theta}:=g^{ ( \alpha_{\theta,2} + r_{\theta,2} \tp ) \vecd_1^\ast - r_{\theta,2} \vecd_3^\ast + ( - \alpha_{\theta,3} + r_{\theta,4} \tp ) \vecd_4^\ast - r_{\theta,4} \vecd_6^\ast }.
        \]
\item
Output
$\KU_\tp:=\{(\tp,\theta,\sKp_{\tp,\theta})\}_{\theta\in\KUNodes(\BT,
\RLp, \tp)}$.
\end{IEEEenumerate}

\item $\DecKeyGen(\SK_{\idp},\KU_\tp)$ On input a private secret key $\SK_{\idp}:=\{(i,\sKp_{\idp,i})\}_{i \in \mathsf{I}}$, $\KU_\tp:=\{(j,\sKp_{\tp,j})\}_{j\in \mathsf{J}}$ for some set of nodes $\mathsf{I},\mathsf{J}$, it runs the following steps:
\begin{IEEEenumerate}
\item $\forall (i,\sKp_{\idp,i})\in \SK_{\idp},(j,\sKp_{\tp,j})\in\KU_\tp$, if $\exists (i, j)$  s.t.\ $i=j$ then $\DK_{\idp,\tp}\leftarrow (\sKp_{\idp,i},\sKp_{\tp,j})$;
else (if $\SK_{\idp}$ and $\KU_\tp$ do not have any node in common)
$\DK_{\idp,\tp}\leftarrow \bot$.
\item Output $\DK_{\idp,\tp}$.
\end{IEEEenumerate}

\item $\Enc(\PP,\idp,\tp,\msgp)$ On input the public parameters $\PP$, an identity $\idp$, a time $\tp\in \Z_q^n$, and a message $\msgp$, it picks $z_1,z_2\getsr\Z_q$ and forms the ciphertext as
\begin{align*}
&\CT_{\idp,\tp}:=\biggl\{\sCp:=\msgp\cdot (g_T^\alpha)^{z_1}, \sCp_0:=g^{ z_1 ( \vecd_1 + \idp \vecd_2+ \tp \vecd_3 ) + z_2 ( \vecd_4 + \idp \vecd_5+ \tp \vecd_6 ) }
\biggr\}.
\end{align*}

\item $\Dec(\PP,\DK_{\idp,\tp},\CT_{\idp,\tp})$  On input the public parameters $\PP$, a decryption key $\DK_{\idp,\tp}:= (\sKp_{\idp,\theta},\sKp_{\tp,\theta})$, and a ciphertext $\CT_{\idp,\tp}:= (\sCp, \sCp_0)$, it computes the message as
        \[
            \msgp:=\sCp/\left(e(\sCp_0,\sKp_{\idp,\theta})\cdot e(\sCp_0,\sKp_{\tp,\theta})\right).
        \]

\item $\KeyRev(\idp,\tp,\RLp,\STp)$  On input an identity $\idp$, a time $\tp$, the revocation list $\RLp$, and the state $\STp$, the algorithm adds $(\idp, \tp)$ to $\RLp$ for all nodes $\nu$ associated with identity $\idp$ and returns $\RLp$.
\end{IEEEitemize}

\noindent \textit{Correctness.} Observe that
\begin{align*}
&e(\sCp_0,\sKp_{\idp,\theta})\\
=\quad &e(g^{z_1 ( \vecd_1 + \idp \vecd_2 + \tp \vecd_3 ) + z_2 ( \vecd_4 + \idp \vecd_5 + \tp \vecd_6 )},g^{( \alpha_{\theta,1} + r_{\theta,1} \idp ) \vecd_1^\ast - r_{\theta,1} \vecd_2^\ast + ( \alpha_{\theta,3} + r_{\theta,3} \idp ) \vecd_4^\ast - r_{\theta,3} \vecd_5^\ast})\\
=\quad &e(g,g)^{\alpha_{\theta,1} z_1\vecd_1\cdot \vecd_1^\ast + \alpha_{\theta,3} z_2 \vecd_4\cdot \vecd_4^\ast}.
\end{align*}
Similarly, $e(\sCp_0,\sKp_{\tp,\theta})=e(g,g)^{\alpha_{\theta,2}
z_1\vecd_1\cdot \vecd_1^\ast - \alpha_{\theta,3} z_2\vecd_4\cdot
\vecd_4^\ast}$. Then
\begin{align*}
&e(\sCp_1,\sKp_{\idp,\theta})\cdot e(\sCp_1,\sKp_{\tp,\theta})\\
=\quad &e(g,g)^{\alpha_{\theta,1} z_1\vecd_1\cdot \vecd_1^\ast + \alpha_{\theta,3} z_2 \vecd_4\cdot \vecd_4^\ast}\cdot e(g,g)^{\alpha_{\theta,2}
z_1\vecd_1\cdot \vecd_1^\ast - \alpha_{\theta,3} z_2\vecd_4\cdot \vecd_4^\ast}\\
=\quad &g_T^{(\alpha_{\theta,1} + \alpha_{\theta,2})z_1}\\
=\quad &(g_T^{\alpha})^{z_1}.
\end{align*}

\subsection{Proof of Security}
We show the RIBE scheme is secure by the following theorem, the
proof techniques are essentially the same as those for
Theorem \ref{thm:security_ribe} except that we use the DLIN-based
Subspace assumption of~\cite{Lewko12}.
\begin{mythm}\label{thm:security_ribe_DLIN}
The RIBE scheme is adaptively secure and anonymous under the DLIN
assumption. More precisely, for any adversary $\AdvA$ against the
RIBE scheme, there exist probabilistic algorithms
\begin{align*}
&\AdvB_0,\\
&\{\AdvB_{\kappa_1,\kappa_2}\}_{\kappa_1=1,\ldots,\qnop,\kappa_2=1,\ldots,\lceil
\log{N_{max}}\rceil},\\
&\{\AdvB_{\kappa_1,\kappa_2}\}_{\kappa_1=\qnop+1,\ldots,\qnop+\qntp+1,\kappa_2=1,\ldots,N_{max}},\\
&\{\AdvB_{\qnop+\qntp+1,\kappa_2}\}_{\kappa_2=1,\ldots,4N_{max}}
\end{align*}
whose running times are essentially the same as that of $\AdvA$,
such that
\begin{align*}
&\Adv_{\AdvA}^{\RIBEp}(\lambda) \leq (\qnop \qntp)^2\cdot \biggl( \Adv_{\AdvB_0}^{\DLINp}(\lambda) + \sum_{\kappa_1=1}^{\qnop}\sum_{\kappa_2=1}^{\lceil \log{N_{max}}\rceil}\Adv_{\AdvB_{\kappa_1,\kappa_2}}^{\DLINp}(\lambda) + \sum_{\kappa_1=\qnop+1}^{\qntp}\sum_{\kappa_2=1}^{N_{max}}\Adv_{\AdvB_{\kappa_1,\kappa_2}}^{\DLINp}(\lambda) \biggr.\\
&\biggl.\quad\quad\quad\quad\quad\quad + \sum_{\kappa_2=1}^{4N_{max}}\Adv_{\AdvB_{\qnop+\qntp+1,\kappa_2}}^{\DLINp}(\lambda) + \frac{6(\qnop \lceil \log{N_{max}}\rceil +\qntp N_{max} ) + 32 N_{max} +6}{q} \biggr)
\end{align*}
where $\qnop,\qntp\geq 4$ are the maximum number of $\AdvA$'s private key
and key update queries respectively.
\end{mythm}

\section{Conclusions}\label{conclusion}

In this paper, we presented two efficient RIBE schemes under the
SXDH and the DLIN assumptions, respectively, which overcome the existing problem of increasing sizes of public parameters. In comparison
with the existing schemes of~\cite{BoldyrevaGK08,LibertV09:ribe},
our RIBE schemes are adaptively secure, anonymous and have constant-size public
parameters, although they have larger sizes of keys and ciphertexts.
Our RIBE schemes can be extended very naturally to obtain revocable IPE schemes with weakly attribute-hiding~\cite{OkamotoT09,OkamotoT10f}. Also our techniques can be applied to a more generally setting, for example, the ABE schemes of~\cite{OkamotoT10f} to obtain adaptively secure revocable ABE schemes.





\appendices

\end{document}